\documentclass{IEEEtran}

\newcommand{\norm}[1]{\lVert#1\rVert}

\usepackage{balance}                
\usepackage{dirtytalk} 
\usepackage{siunitx}						
\usepackage[official]{eurosym} 

\usepackage{soul}
\usepackage{makecell}

\makeatletter
\newcommand*{\bigs}[1]{{\hbox{$\left#1\vbox to10\p@{}\right.\n@space$}}}
\makeatother

\makeatletter
\newcommand*{\biggs}[1]{{\hbox{$\left#1\vbox to17\p@{}\right.\n@space$}}}
\makeatother

\usepackage{blindtext}

\usepackage{tikz}  
\usetikzlibrary{decorations.pathreplacing,calligraphy}
\usepackage[framemethod=tikz]{mdframed}
\usetikzlibrary{shapes.geometric, arrows, arrows.meta}
\usepackage{subcaption}
\usepackage{graphicx}
\usepackage{float}
\usepackage{booktabs}
\usepackage{multirow}
\usepackage{dblfloatfix}
\DeclareSIUnit[quantity-product = ]\percent{\char`\%}
\usepackage{comment}
\usetikzlibrary{decorations.text}
\usetikzlibrary{shapes.geometric, arrows}
\usetikzlibrary{positioning}
\usepackage{adjustbox}

\usetikzlibrary{intersections}

\def\,{\mskip\thinmuskip} \def\!{\mskip-\thinmuskip}
\def\,{\mskip\thickmuskip} \def\;{\mskip+\thickmuskip}

\usepackage{cite}
\usepackage{amsmath,amssymb,amsfonts}
\usepackage{algorithmic}
\usepackage{graphicx,color}
\usepackage{textcomp}
\usepackage{xcolor}
\usepackage{hyperref}
\hypersetup{hidelinks=true}
\usepackage{algorithm,algorithmic}
\def\BibTeX{{\rm B\kern-.05em{\sc i\kern-.025em b}\kern-.08em
    T\kern-.1667em\lower.7ex\hbox{E}\kern-.125emX}}

\AtBeginDocument{\definecolor{tmlcncolor}{cmyk}{0.93,0.59,0.15,0.02}\definecolor{NavyBlue}{RGB}{0,86,125}}

\usepackage{bm}
\usepackage{comment}

\usepackage{pifont} 
\makeatletter
\newcommand*\bigcdot{\mathpalette\bigcdot@{.5}}
\newcommand*\bigcdot@[2]{\mathbin{\vcenter{\hbox{\scalebox{#2}{$\m@th#1\bullet$}}}}}
\makeatother
\usepackage{tikz}  
\usepackage{url}

\definecolor{mygreen}{RGB}{127,201,127}
\definecolor{mypurple}{RGB}{190,174,212}
\definecolor{myorange}{RGB}{253,192,134}
\definecolor{myblue}{RGB}{72,213,210}

\usepackage{color}
\definecolor{light}{rgb}{0.5, 0.5, 0.5}

\definecolor{answerblue}{rgb}{0.21,0.37,0.57}
\newcommand\Change[1]{#1}

\usepackage{booktabs, makecell, siunitx, adjustbox}

\begin{document}

\title{MambAttention: Mamba with Multi-Head Attention for Generalizable Single-Channel Speech Enhancement}

\author{Nikolai Lund Kühne,
\IEEEmembership{Graduate Student Member, IEEE}, Jesper Jensen, \IEEEmembership{Member, IEEE},

Jan Østergaard, \IEEEmembership{Senior Member, IEEE}, and Zheng-Hua Tan, \IEEEmembership{Senior Member, IEEE}
\thanks{This work was supported by DeiC by enabling access to the LUMI supercomputer (g.a. DeiC-AAU-N5-2025126-``Exploring new neural architectures for deep learning based speech enhancement'').}
\thanks{This work is partially supported by the William Demant Fonden via the Centre for Acoustic Signal Processing Research (CASPR).}
\thanks{Nikolai Lund Kühne, Jan Østergaard, and Zheng-Hua Tan are with the Department of Electronic Systems, Aalborg University, 9220, Denmark  (e-mail: nlk@es.aau.dk; jo@es.aau.dk; zt@es.aau.dk). Zheng-Hua Tan is also with the Pioneer Centre for AI, 1350, Denmark.}
\thanks{Jesper Jensen is with the Department of Electronic Systems, Aalborg University, 9220, Denmark, and also with Oticon A/S, 2765, Denmark (e-mail: jesj@demant.com).}
}

\maketitle

\begin{abstract}
With the advent of new sequence models like Mamba and xLSTM, several studies have shown that these models match or outperform the state-of-the-art in single-channel speech enhancement and self-supervised audio representation learning. However, prior research has demonstrated that sequence models like LSTM and Mamba tend to overfit to the training set. To address this issue, previous works have shown that adding self-attention to LSTMs substantially improves generalization performance for single-channel speech enhancement.
Nevertheless, neither the concept of hybrid Mamba and time-frequency attention models nor their generalization performance  have been explored for speech enhancement. In this paper, we propose a novel hybrid architecture, MambAttention, which combines Mamba and shared time- and frequency-multi-head attention modules for generalizable single-channel speech enhancement. To train our model, we introduce VB-DemandEx, a dataset inspired by VoiceBank+Demand but with more challenging noise types and lower signal-to-noise ratios. Trained on VB-DemandEx, MambAttention significantly outperforms existing state-of-the-art discriminative LSTM-, xLSTM-, Mamba-, and Conformer-based systems of similar complexity across all reported metrics on two out-of-domain datasets: DNS 2020 without reverberation and EARS-WHAM\_v2. MambAttention also matches or outperforms generative models such as diffusion models in generalization performance while being competitive with language model baselines. Ablation studies highlight the importance of weight sharing between time- and frequency-multi-head attention modules for generalization performance. Finally, we explore integrating the shared time- and frequency-multi-head attention modules with LSTM and xLSTM, which yields a notable performance improvement on the out-of-domain datasets. However, MambAttention remains superior for cross-corpus generalization across all reported evaluation metrics.
\end{abstract}


\section{INTRODUCTION}
\IEEEPARstart{S}{peech} enhancement aims to improve the speech intelligibility and speech quality of noisy speech signals, by removing background noise and recovering the desired speech signal. It is a widely studied subject, since speech enhancement is both challenging and has a wide array of applications such as hearing assistive devices, mobile communication devices, speech recognition systems, and speaker verification systems. 

Over the last decade, research on the single-microphone setting, also known as single-channel speech enhancement, developed from using classical signal-processing techniques such as Kalman filtering \cite{paliwal1987speech} and Minimum Mean Square Error Short-Time Spectral Amplitude estimation \cite{ephraim1984speech} to using deep neural networks (DNNs) \cite{xu2014regression, kolbaek2016speech}. As the field of deep learning evolves, new neural architectures emerge. This has led to a large selection of deep learning-based single-channel speech enhancement systems using a variety of neural architectures such as deep denoising autoencoders \cite{lu2013speech}, recurrent neural networks (RNNs) and Long Short-Term Memory (LSTMs) networks \cite{tesch22_interspeech, hao2021fullsubnet}, convolutional neural networks \cite{fu2016snr,pandey2019new,kolbaek2020loss, yan2025lisennet}, diffusion models \cite{lu2022conditional, richter2023speech, richter2024causal, gonzalez2024investigating}, generative adversarial networks \cite{michelsanti17_interspeech, pascual17_interspeech, fu2019metricgan, fu2021metricgan+, abdulatif2024cmgan}, and state-space models (SSMs) \cite{sun2024dual,ku23_interspeech,du2024spiking}. With the advent of the Transformer \cite{vaswani2017attention}, attention-based speech enhancement systems have achieved state-of-the-art on several datasets \cite{mp-senet, lu2023explicit}.

However, scaled dot-product attention-based models like Transformers and Conformers \cite{gulati20_interspeech} scale poorly with sequence length and require large training datasets \cite{deoliveira22_interspeech, gong21b_interspeech}. Hence, recent works have focused on proposing sequence models with linear scalability with respect to sequence length such as Mamba \cite{gu2024mamba} and Extended Long Short-Term Memory (xLSTM) \cite{xlstm}. Mamba and xLSTM have already shown great promise in natural language processing (NLP) \cite{gu2024mamba, xlstm}, computer vision \cite{vim, visionxlstm}, and self-supervised audio representation learning \cite{audiomamba, audioxlstm}. Additionally, Mamba and xLSTM have recently been shown to match or outperform state-of-the-art speech enhancement systems \cite{semamba, xlstm-senet}. Interestingly, \cite{xlstm-senet} also found that a correctly configured LSTM-based model can actually match Conformer-, Mamba-, and xLSTM-based systems on the \textit{VoiceBank+Demand} dataset \cite{voicebank, demand}. However, many papers reporting state-of-the-art performance often only evaluate in-domain speech enhancement performance. Arguably, in-domain speech enhancement performance may not be representative of performance in real-world environments, where speech and noise signals may vary significantly from the training data. For this reason, we focus on developing a speech enhancement algorithm that yields superior cross-corpus generalization performance. 

In this paper, we propose MambAttention (\autoref{fig: TF Mamba-Attention}), which is a novel hybrid Mamba and multi-head attention (MHA) model for generalizable single-channel speech enhancement. MambAttention comprises both time- and frequency-Mamba modules (T- and F-Mamba) as well as time- and frequency-MHA modules (T- and F-MHA). More importantly, our MambAttention model shares weights between the T- and F-MHA modules within each layer. By sharing the weights between the MHA modules in each layer, our MambAttention model aligns global time- and frequency content by jointly attending to both temporal and spectral features. We find that this substantially improves generalization across recording conditions, speakers and noise types. To the best of our knowledge, MambAttention is the first model to combine Mamba with MHA across time and frequency for speech enhancement.

While \textit{VoiceBank+Demand} \cite{voicebank,demand} is a widely-used benchmark for single-channel speech enhancement, the test set contains neither babble noise nor speech-shaped noise (SSN) and the signal-to-noise ratios (SNRs) of both training and test files are not lower than \SI{0}{dB}. This means speech enhancement systems trained and tested only on \textit{VoiceBank+Demand} are neither exposed to nor evaluated in difficult listening conditions. Motivated by this, we propose a new benchmark: \textit{VoiceBank+Demand Extended} (\textit{VB-DemandEx}). Compared to the \textit{VoiceBank+Demand} dataset, our \textit{VB-DemandEx} comprises much lower SNRs and a larger variety of noise types.

To evaluate the performance of MambAttention, we train our models on \textit{VB-DemandEx}, and test on the in-domain \textit{VB-DemandEx} test set as well as out-of-domain test sets from \textit{Deep Noise Suppression  Challenge 2020} \cite{reddy20_interspeech} (\textit{DNS 2020}) and \textit{EARS-WHAM\_v2} \cite{richter24_interspeech}. Results show that our MambAttention model significantly outperforms state-of-the-art discriminative LSTM-, xLSTM-, Mamba-, and Conformer-based systems on the out-of-domain test sets across all reported evaluation metrics, while maintaining state-of-the-art performance on the in-domain test set. In addition, our MambAttention model matches or outperforms diffusion models and is competitive with language model (LM) baselines on the \textit{DNS 2020} real-recordings test set, and the test set without reverberation. Despite not being trained for reverberation, MambAttention still rivals some diffusion baselines on this task. Ablation studies on our proposed MambAttention model shows that the weight sharing mechanism positively impacts generalization performance. Additionally, we find that placing the MHA modules before the Mamba blocks significantly improves generalization performance.
Furthermore, we find that augmenting existing LSTM- and xLSTM-based speech enhancement systems with our proposed MHA modules also greatly improves generalization performance; however, our MambAttention model remains superior across all reported evaluation metrics. 

To gain further insights, we visualize the latent features of our MambAttention model, and the LSTM-, xLSTM-, Mamba-, and Conformer-based models. 
Our findings suggest that MHA encourages the model to learn dataset-invariant representations. 
Finally, our MambAttention model shows superior scalability with respect to dataset size compared to discriminative LSTM, xLSTM, Mamba, and Conformer baseline, when trained on the large-scale \textit{DNS 2020} dataset.

Our major contributions are summarized as follows:
\begin{itemize}
    \item We propose a novel state-of-the-art hybrid MambAttention model combining Mamba and MHA for generalizable single-channel speech enhancement.
    \item We demonstrate that MambAttention outperforms discriminative baselines and matches or surpasses generative diffusion models and is competitive with large LMs for generalizable speech enhancement.
    \item We demonstrate that weight sharing between T- and F-MHA modules in our MambAttention model contributes to its state-of-art generalization performance.
    \item We show that combining our shared T- and F-MHA modules with LSTM- and xLSTM-based models significantly improves their generalization performance.
    \item We propose the \textit{VB-DemandEx} benchmark, which is inspired by \textit{VoiceBank+Demand}, but features substantially lower SNRs and more noise types.
\end{itemize}
Code for training and inference, audio samples, model weights, and the proposed dataset are publicly available.\footnote{\url{https://github.com/NikolaiKyhne/MambAttention}}

\section{Related Works}

\subsection{Generalization Performance of Sequence Models}
Prior works have found that LSTMs overfit to the training dataset for automatic speech recognition in RNN-Transducers, which results in poor generalization performance \cite{chiu2021rnn,kim22f_interspeech}. To remedy this, \cite{chiu2021rnn} combines multiple regularization techniques during training, and uses dynamic overlapping inference by segmenting long utterances into multiple fixed-length segments, which are decoded independently. 
On the other hand, \cite{kim22f_interspeech} proposes using sparse self-attention layers in the Conformer RNN-Transducer. Additionally, they hypothesize that during inference on long utterances, unseen linguistic context accumulate excessively in the hidden state of the LSTM, which can be mitigated by resetting the hidden states at silent segments. Similarly, \cite{long2024dgmamba} hypothesizes that domain-specific information can potentially be
accumulated or even amplified in hidden states during propagation in the Mamba architecture, thus resulting in worse generalization performance. To mitigate this, they propose hidden state suppressing, which reduces the gap between hidden states across domains, and find  that this improves generalization performance \cite{long2024dgmamba}. 

In \cite{pandey2020cross}, it was concluded that poor generalization performance of speech enhancement DNNs often stems from different recording conditions between datasets. It is revealed that the content of a corpus is more important than its size for generalization performance. To overcome poor enhancement performance on unseen datasets, self-attending RNNs were proposed in \cite{pandey2022self}. By adding self-attention to RNNs, cross-corpus generalization was significantly improved. In \cite{gonzalez2023assessing}, a generalization assessment framework that accounts for a potential change in the difficulty of the speech enhancement task across datasets is presented. Using this framework, it was found that performance degrades the most in speech mismatches. Notably, it was revealed that while the most recent models displayed the best in-domain speech enhancement performance, their out-of-domain speech enhancement performance was beaten by the older systems \cite{gonzalez2023assessing}.
\begin{figure*}[!thb]
    \centering
    \input{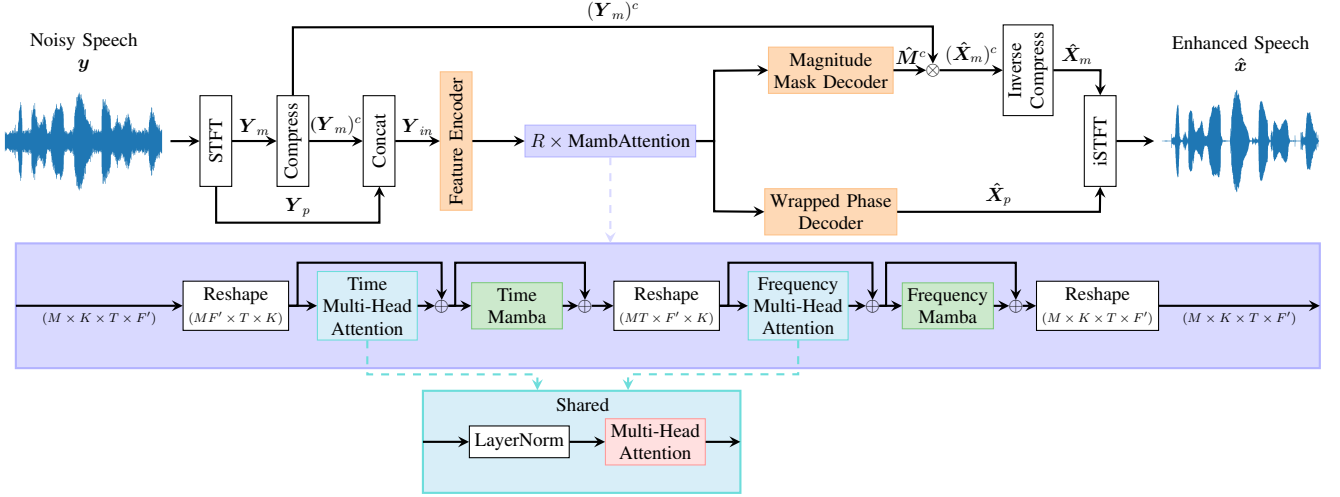}
    \caption{Overall structure of our proposed MambAttention model. $M$, $K$, $T$, and $F'$ represent the batch size, the number of channels, the number of time frames, and the number of frequency bins, respectively.}
    \label{fig: TF Mamba-Attention}
\end{figure*}
\subsection{Generalization Performance of Generative Models}
While discriminative speech enhancement models learn a direct mapping between noisy and clean speech, generative models are trained to learn the distribution of the clean speech signals as a prior for generation of enhanced speech signals \cite{lemercier2025diffusion}. Several works have demonstrated that generative models such as diffusion models exhibit remarkable generalization performance for speech enhancement \cite{lu2022conditional, SGMSE, lemercier2023storm, lemercier2025diffusion}. However, diffusion models are computationally expensive during inference, as they require running the backbone neural network for each reverse diffusion step \cite{lemercier2023storm}. Nevertheless, recent work has demonstrated the potential of causal diffusion models for generalized speech enhancement \cite{richter2024causal}.

Besides diffusion models, SELM was proposed in \cite{wang2024selm} demonstrating the potential of LMs for denoising and dereverberation. By teaching the LM to predict the target probability distribution given the input probability distribution, both denoising and dereverberation are learnt through a cross-entropy loss function \cite{wang2024selm}. Several papers have extended the idea of using LMs for speech enhancement, improving both the performance of the models and the amount of tasks the speech LMs can handle \cite{masksr,anyenhance,yao2025gense,llase}. However, training speech LMs for speech enhancement often requires several thousand hours of training data \cite{anyenhance,yao2025gense, llase}, which comprises multiple clean-speech and noise datasets covering all tasks. In addition, the model size of the LMs for speech enhancement is often more than a hundred times larger than specialized models, leading to high computational complexity. While speech LMs have demonstrated impressive speech enhancement performance, some models like UniAudio \cite{yang2024uniaudio} only handle one distortion at a time and requires task-specific fine-tuning. Recent works, however, have proposed a unified approch allowing the speech LM to perform multiple tasks at the same time \cite{masksr, anyenhance,llase}.

\subsection{Hybrid Sequence Models}
Combining Mamba blocks with Transformer or attention blocks has already shown great promise in NLP \cite{lenz2025jamba, rensamba}, where the proposed models match or outperform state-of-the-art models on multiple benchmarks. However, the large token contexts used in NLP prevent the use of self-attention and MHA. Similarly, for automatic speech recognition, combining a Transformer encoder with an SSM results in state-of-the-art performance \cite{fathullah23_interspeech}. 
Beyond NLP and automatic speech recognition, a hybrid Mamba and MHA model DeFT-Mamba \cite{deftmamba} was recently proposed for universal audio separation in multichannel polyphonic scenarios. By combining a gated convolution block with a hybrid Mamba and MHA block across frequency- and time-dimensions, DeFT-Mamba demonstrates state-of-the-art performance in multichannel separation. Finally, in \cite{sui2024tramba}, Transformer and Mamba layers were combined in a U-Net architecture called TRAMBA. TRAMBA outperforms existing models in practical speech super resolution and enhancement on mobile and wearable platforms in a self-supervised setting.

As opposed to \cite{lenz2025jamba, rensamba, fathullah23_interspeech}, the MHA modules in MambAttention are placed before their respective Mamba blocks. Additionally, in contrast to DeFT-Mamba \cite{deftmamba}, our MambAttention model processes the encoded input along the time-dimension before the frequency-dimension. As we will demonstrate, this ordering yields superior out-of-domain speech enhancement performance. Moreover, while DeFT-Mamba employs a gated convolution block, which shifts features by one frequency bin or one time frame before applying the MHA and Mamba blocks to process across frequency or time, our model does not. Finally, as opposed to most existing hybrid attention models \cite{pandey2022self,lenz2025jamba,rensamba, fathullah23_interspeech, sui2024tramba, deftmamba}, MambAttention shares the weights of the T- and F-MHA modules in each layer.
\section{PROPOSED METHOD}\label{sec:method}
\subsection{State-Space Models and Mamba}
Structured SSMs \cite{gu2022efficiently} and Mamba \cite{gu2024mamba} are a family of sequence-to-sequence models inspired by continuous linear time-invariant systems. SSMs map an input $x(t)\in\mathbb{R}$ to an output $y(t)\in\mathbb{R}$ through a latent state $\bm{h}(t)\in\mathbb{R}^{N\times 1}$ via an evolution parameter $\bm{A}\in\mathbb{R}^{N\times N}$, and projection parameters $\bm{B}\in\mathbb{R}^{N\times 1}$ and $\bm{C}\in\mathbb{R}^{1\times N}$:
\begin{align}
    \bm{h}'(t) &= \bm{A}\bm{h}(t) + \bm{B}x(t),\label{eq:LTISSM1}\\
    y(t) &= \bm{C}\bm{h}(t),\label{eq:LTISSM2}
\end{align}
where $\bm{h}'(t)\in\mathbb{R}^{N\times 1}$. To make SSMs applicable in deep neural networks, a time-scale parameter $\Delta\in\mathbb{R}$ is introduced to transform $\bm{A}$ and $\bm{B}$ into their discrete-time counterparts $\overline{\bm{A}}\in\mathbb{R}^{N\times N}$ and $\overline{\bm{B}}\in\mathbb{R}^{N\times 1}$ via a zero-order hold \cite{han2024demystify}:
\begin{align}
    \overline{\bm{A}} &= \mathrm{exp}{(\Delta \bm{A})},\label{eq:LTIZOH1}\\
    \overline{\bm{B}} &= (\Delta \bm{A})^{-1}(\mathrm{exp}(\Delta\bm{A}) -\bm{I})\cdot \Delta\bm{B}\approx \Delta\bm{B}. \label{eq:LTIZOH2}
\end{align}
The approximation in \eqref{eq:LTIZOH2} becomes increasingly accurate when $\Delta$ is small.
Thus, the discrete-time versions of \eqref{eq:LTISSM1} and \eqref{eq:LTISSM2} become \cite{gu2024mamba}:
\begin{align}
    \bm{h}_i &= \overline{\bm{A}}\bm{h}_{i-1} + \overline{\bm{B}}x_i,\label{eq:LTISSM3}\\
    y_i &= \bm{C}\bm{h}_i,\label{eq:LTISSM4}
\end{align}
where the subscript $i$ is the discrete-time index.
Mamba improves upon structured SSMs, by making $\bm{B},\bm{C},\Delta$ functions of the input, resulting in the input-dependent parameters $\bm{B}_i\in\mathbb{R}^{N\times 1}$, $\bm{C}_i\in\mathbb{R}^{1\times N}$, and $\Delta_i\in\mathbb{R}$. Consequently, the discretized $\overline{\bm{A}}_i=\mathrm{exp}(\Delta_i\bm{A})$, $\overline{\bm{B}}_i=\Delta_i\bm{B}_i$ also become input-dependent. Additionally, Mamba sets $\bm{A}$ and $\overline{\bm{A}}_i$ as diagonal; hence,  defining the vector composed of diagonal elements of $\overline{\bm{A}}_i$ as $\Tilde{\bm{A}}_i=\mathrm{diag}(\overline{\bm{A}}_i)\in\mathbb{R}^{N\times 1}$ results in $\overline{\bm{A}}_i\bm{h}_{i-1}=\Tilde{\bm{A}}_i\odot \bm{h}_{i-1}$, where $\odot$ is element-wise multiplication. Moreover, we can write $\overline{\bm{B}}_ix_i=\Delta_i\bm{B}_ix_i=\bm{B}_i(\Delta_i\odot x_i)$ \cite{han2024demystify}. Thus, \eqref{eq:LTISSM3} and $\eqref{eq:LTISSM4}$ become:
\begin{align}
     \bm{h}_i &= \Tilde{\bm{A}}_i\odot \bm{h}_{i-1} + \bm{B}_i (\Delta_i\odot x_i),\label{eq:LTISSM5}\\
    y_i &= \bm{C}_i\bm{h}_{i},\label{eq:LTISSM6}
\end{align}
where $x_i,y_i\in\mathbb{R}$, and $\bm{h}_i\in\mathbb{R}^{N\times 1}$.

Finally, to operate over an input $\bm{X}=[\bm{x}_1, \bm{x}_2, \dots,\bm{x}_L]^{\top}\in\mathbb{R}^{L\times K}$, where each $\bm{x}_i\in\mathbb{R}^{1\times K}$, Mamba applies \eqref{eq:LTISSM5} and \eqref{eq:LTISSM6} independently to each channel $K$. Thus, the formulation of Mamba becomes \cite{han2024demystify}:
\begin{align}
    \bm{h}_i &= \Tilde{\bm{A}}_i\odot \bm{h}_{i-1} + \bm{B}_i(\bm{\Delta}_i\odot \bm{x}_i),\\
    \bm{y}_i &= \bm{C}_i\bm{h}_{i},
\end{align}
where $\bm{\Delta}_i\in\mathbb{R}^{1\times K}$, $\bm{B}_i\in\mathbb{R}^{N\times 1}$, $\bm{C}_i\in\mathbb{R}^{1\times N}$ are derived from the input, $\Tilde{\bm{A}}_i,\bm{h}_{i}\in\mathbb{R}^{N\times K}$, and $\bm{y}_i\in\mathbb{R}^{1\times K}$. In Mamba, the parameters $\bm{B}\in\mathbb{R}^{N\times L}$, $\bm{C}\in\mathbb{R}^{L\times N}$, and $\bm{\Delta}\in\mathbb{R}^{L\times K}$ are learned though projections $\bm{B}=(\bm{X}\bm{W}_B)^{\top}$, $\bm{C}=\bm{X}\bm{W}_C$, and $\bm{\Delta}=\mathrm{Softplus}(\bm{X}\bm{W}_1\bm{W}_2)$, where $\bm{W}_B,\bm{W}_C\in\mathbb{R}^{K\times N}$, $\bm{W}_1\in\mathbb{R}^{K\times K_0}$, and $\bm{W}_2\in\mathbb{R}^{K_0\times K}$ are learnable weight matrices \cite{han2024demystify}, $\bm{X}$ is the input, and the $\mathrm{Softplus}$ function is defined in \cite{zheng2015improving}.

\subsection{Multi-Head Attention}
Attention can be interpreted as a vector of importance weights assigned to different parts of the input or output based on context derived from learnable feature spaces \cite{vaswani2017attention}. Self-attention is an attention mechanism relating different positions of the same input sequence by assigning different weights to different parts of the input sequence \cite{vaswani2017attention}. Informally, an attention mechanism is a mapping of a query and a set of key-value pairs to an output. The popular scaled dot-product attention mechanism is defined as \cite{vaswani2017attention}:
\begin{align}
        \mathrm{Attention}(\bm{Q},\bm{K},\bm{V}) = \mathrm{Softmax}\left(\frac{\bm{Q}\bm{K}^{\top}}{\sqrt{d_k}}\right)\bm{V},
\end{align}
where $\bm{Q}\in\mathbb{R}^{L\times d_k}$, $\bm{K}\in\mathbb{R}^{L\times d_k}$ and $\bm{V}\in\mathbb{R}^{L\times d_v}$ is the query, key, and value matrix, respectively, and $L$ is the input sequence length. The query, key, and value matrices are learned projections of the input $\bm{X}\in\mathbb{R}^{L \times d_m}$ to their respective dimensions and $d_m$ is the model’s feature dimension. In a single self-attention computation, all information from the input is averaged. However, in the MHA mechanism proposed in \cite{vaswani2017attention}, scaled dot-product attention is computed across $h$ attention heads of dimensionality $\frac{d_m}{h}=d_k=d_v$ in parallel. This allows neural networks utilizing MHA to jointly attend to information from different subspace representations at different positions, meaning that different attention heads capture different information. Given $\bm{Q},\bm{K},\bm{V}\in\mathbb{R}^{L\times d_m}$, the MHA mechanism is given by:
\begin{align}
    \mathrm{MHA}(\bm{Q},\bm{K},\bm{V}) &= \mathrm{Concat}(\mathrm{head}_1,\dots,\mathrm{head}_h)\bm{W}^O,\\
    \mathrm{head}_i &= \mathrm{Attention}(\bm{Q}\bm{W}_i^Q,\bm{K}\bm{W}_i^K,\bm{V}\bm{W}_i^V),
\end{align}
 where $\mathrm{Concat}(\cdot)$ is the concatenation operation, $\bm{W}_i^Q,\bm{W}_i^K\in\mathbb{R}^{d_m\times d_k}$, and $\bm{W}_i^V\in\mathbb{R}^{d_m\times d_v}$ are learnable weight matrices, $\mathrm{head}_i\in\mathbb{R}^{L\times \frac{d_m}{h}}$ denotes the $i$th attention head, and $i\in\{1,2,\dots,h\}$. Finally, $\bm{W}^O\in\mathbb{R}^{hd_v\times d_m}$ is the output weight matrix, learning the contribution of each attention head.

\subsection{MambAttention: Mamba with Multi-Head Attention}
To improve generalization performance for speech enhancement, we propose the MambAttention block, which is a novel \Change{non-causal} architectural component integrating shared MHA modules with Mamba. \autoref{fig: TF Mamba-Attention} illustrates our proposed MambAttention block. Each block jointly models temporal and spectral dependencies, enabling the network to capture complex structures in speech signals. In each MambAttention block, the input $\bm{X}\in\mathbb{R}^{M\times K\times T \times F}$ is first reshaped to $MF\times T \times K$ before applying a Layer Normalization (LN), a T-MHA block and a bidirectional T-Mamba block. Here $M$, $K$, $T$, and $F$ represent the batch size, the number of channels, time frames, and frequency bins, respectively. Subsequently, the output of the T-Mamba block is reshaped to $MT\times F \times K$ before another LN, a F-MHA block, and a bidirectional F-Mamba block is applied. Finally, the output is reshaped back to ${M\times K\times T \times F}$. Mathematically, given an input $\bm{X}$, the forward pass of a MambAttention block is given by:
\begin{align}
    \bm{X}_{\mathrm{Time}} &= \mathrm{reshape}(\bm{X},[M\cdot F,T,K]), \\
    \bm{X}_1 &= \bm{X}_{\mathrm{Time}} + \text{T-MHA}(\mathrm{LN}(\bm{X}_{\mathrm{Time}})), \\
    \bm{X}_2 &= \bm{X}_1 + \text{T-Mamba}(\bm{X}_1), \\
    \bm{X}_{\mathrm{Freq.}} &= \mathrm{reshape}(\bm{X}_2,[M\cdot T,F,K]), \\
    \bm{X}_3 &= \bm{X}_{\mathrm{Freq.}} + \text{F-MHA}(\mathrm{LN}(\bm{X}_{\mathrm{Freq.}})),\\
    \bm{X}_4 &= \bm{X}_3 + \text{F-Mamba}(\bm{X}_3), \\
    \bm{Y} &= \mathrm{reshape}(\bm{X}_4,[M, K, T,F]),
\end{align}
where $\mathrm{reshape}(\mathrm{input}, \mathrm{size})$ reshapes the input to a given size. The T- and F-MHA modules only have one input, since the queries, keys, and values are all derived from the input. We use the T- and F-Mamba blocks from SEMamba \cite{semamba}, hence the output $\bm{X}_\mathrm{out}$ of each T- and F-Mamba block is given by:
\begin{multline}
    \bm{X}_\mathrm{out} = \mathrm{Conv1D}(\mathrm{Concat}(\mathrm{Mamba}(\bm{X}_\mathrm{in}), \\\mathrm{flip}(\mathrm{Mamba}(\mathrm{flip}(\bm{X}_\mathrm{in}))))),
\end{multline}
where $\bm{X}_\mathrm{in}$ is the input to the T- and F-Mamba blocks, and $\mathrm{Mamba}(\cdot)$, $\mathrm{flip}(\cdot)$, and $\mathrm{Conv1D}(\cdot)$ is the unidirectional Mamba, the sequence flipping operation, and the 1-D transposed convolution across either time or frequency.

A key element of our MambAttention block is the use of shared weights between the T- and F-MHA modules within each MambAttention block. This weight sharing mechanism allows each layer of the model to simultaneously attend to both time and frequency content. Importantly, as we shall see, weight sharing substantially improves the model's ability to generalize across recording conditions, speaker, and noise types. Finally, weight sharing minimizes the increase in model size and memory cost from adding the MHA modules, resulting in more efficient training.
\section{EXPERIMENTAL SETUP}\label{sec:experiments}
\subsection{Datasets}
To generate our proposed \textit{VB-DemandEx} dataset, we use the same clean speech data as \textit{VoiceBank+Demand} \cite{voicebank,demand}, but we leave out speakers ``\textit{p282}'' (female) and ``\textit{p287}'' (male) for validation, as the original \textit{VoiceBank+Demand} benchmark contains no validation set. Like \textit{VoiceBank+Demand}, speakers ``\textit{p232}'' and ``\textit{p257}'' are used for the test set, and the remaining 26 speakers are used for training. All audio clips are downsampled to \SI{16}{kHz}. For noise, \Change{we follow \cite{stylianou2025librivad} and} use the \SI{16}{kHz} version of the Demand database \cite{demand}, which comprises 6 noise categories: Domestic, Office, Public, Transportation, Street, and Nature. In each category, there are 3 subsets of noise recordings. Each subset of noise recordings contains 16 audio signals that are 15 minutes long, enumerated from 1 to 16. In \textit{VoiceBank+Demand}, only selected subsets of each noise category are present in the training and test sets. To ensure our speech enhancement systems are trained on a bigger variety of noise types, we adopt a different approach. For each subset of noise recordings, we divide the signals into 3 groups of 1-12, 13-14 and 15-16, and concatenate them into training, validation and test splits, respectively. This ensures no shared realizations of noise types between the training, validation, and test sets.
The subsets of each noise categories are concatenated further in order to create the splits of 6 different noise types (e.g. $\textrm{Public\_training} = \mathrm{Concat}(\mathrm{PSTATION}[1-12], \mathrm{PCAFETER}[1-12], \mathrm{PRESTO}[1-12])$). In addition to the Demand database, we use babble noise and SSN, which is generated using LibriSpeech \cite{panayotov2015librispeech} as a source, in the training, validation, and test sets. Babble noise is produced by averaging the waveforms of 6 energy-standardized speech signals with silence detected by rVAD \cite{tan2020rvad} being removed, resulting in a single babble signal. To create the SSN, audio signals from selected speakers undergo a 12th order Linear Predictive Coding analysis, yielding coefficients for an all-pole filter that is applied to Gaussian white noise. The train, validation, and test set in \textit{VB-DemandEx}, are generated by sequentially mixing the clean speech and noise at 7 segmental SNRs (SSNRs) \cite{hansen1998effective} ([$-10, -5, 0, 5, 10, 15, 20$] dB), using the officially provided script from the \textit{DNS 2020} dataset \cite{reddy20_interspeech}. This ensures a uniform distribution of both noise types and SNRs. This results in 10,842 audio clips for training, 730 for validation, and 826 for testing.

Besides training and testing on our proposed \textit{VB-DemandEx} dataset, we also train and test our models on the large-scale \textit{DNS 2020} dataset \cite{reddy20_interspeech}. This dataset contains 500 hours of clean audio clips from
2,150 speakers and over 180 hours of noise audio clips. Since the original \textit{DNS 2020} dataset contains no validation set, we set aside female speakers ``\textit{reader\_01326}'', ``\textit{reader\_06709}'', ``\textit{reader\_08788}'' and male speakers ``\textit{reader\_01105}'', ``\textit{reader\_05375}'', ``\textit{reader\_11980}'', as well as a suitable amount of noise audio clips for validation. Following the officially provided script, the remaining clean and noisy audio clips are used to generate 3,000 hours of noisy-clean pairs of audio clips with SSNRs ranging from \SI{-5}{dB} to \SI{15}{dB} for training, resulting in \SI{1.08}{M} \SI{10}{s} noisy audio clips. We generate the validation set using the same script resulting in 315 \SI{10}{s} audio clips with SSNRs between \SI{-5}{dB} and \SI{15}{dB}, ensuring a uniform distribution of noise types and SNRs. For evaluation, we use both the \textit{DNS 2020} test set with reverberation (reverb), without reverb, and the real-recordings test set. The test sets with and without reverb each contain 150 noisy-clean pairs generated from audio clips spoken by 20 speakers. The real-recordings test set contains 300 audio clips without clean references.

We also test the generalization performance of our models on the \SI{16}{kHz} version of the \textit{EARS-WHAM\_v2} dataset \cite{richter24_interspeech,wichern19_interspeech}, which comprises clean audio clips from 107 speakers. The clean speech, which is recorded in an anechoic chamber, covers a large variety of speaking styles, including reading tasks in 7 different reading styles, emotional reading and freeform speech in 22 different emotions, as well as conversational speech \cite{richter24_interspeech}. Speakers \textit{p001} to \textit{p099} are used for training, \textit{p100} and \textit{p101} are used for validation, and \textit{p102} to \textit{p107} are used for testing. Using the officially provided script \cite{richter24_interspeech}, the clean speech is mixed with real noise recordings from the \textit{WHAM!} dataset \cite{wichern19_interspeech} at SNRs randomly sampled between [-2.5, 17.5] dB. The SNRs are computed using loudness K-weighted relative to full scale standardized in ITU-R BS.1770 \cite{ITU-BS1770-5}. This results in 32,485 noisy-clean pairs for training, 632 noisy-clean pairs for validation, and 886 noisy-clean pairs for testing. The \textit{EARS-WHAM\_v2} test set contains utterances up to \SI{29}{s} in length. Hence, to avoid running out of memory, we perform inference on non-overlapping chunks of \SI{10}{s} audio clips for all models on the \textit{EARS-WHAM\_v2} test set.

\subsection{Model Overview}
We integrate our MambAttention block into the widely-used state-of-the-art dual-path system used in MP-SENet \cite{mp-senet} (Conformer), SEMamba \cite{semamba} (Mamba), and xLSTM-SENet \cite{xlstm-senet} (xLSTM). 
Thus, all pre-processing, feature encoding, final decoding, training hyperparameters, and loss functions are equivalent for all discriminative models trained and compared in this paper.

\autoref{fig: TF Mamba-Attention} illustrates the overall architecture of our MambAttention model. A complex spectrogram of the noisy speech waveform $\bm{y}\in\mathbb{R}^D$ is computed via an STFT. The input to the feature encoder $\bm{Y}_{in}\in\mathbb{R}^{T\times F\times 2}$ is the compressed magnitude spectrum $(\bm{Y}_m)^c\in\mathbb{R}^{T\times F}$ extracted via power-law compression \cite{powerlawcompression} concatenated with the wrapped phase spectrum $\bm{Y}_p\in\mathbb{R}^{T\times F}$. The feature encoder contains two convolution blocks sandwiching a dilated DenseNet \cite{densenet}. Each convolution block consists of a 2D convolutional layer, an instance normalization, and a PReLU activation \cite{he2015delving}. The feature encoder increases the number of input channels from $2$ to $K$ and halves the frequency dimension from $F$ to $F'=F/2$.

The output of the feature encoder is then processed by $R$ MambAttention blocks. It is subsequently fed into the magnitude mask decoder and the wrapped phase decoder that predicts the clean compressed magnitude mask $\bm{M}^c = (\bm{X}_m/\bm{Y}_m)^c\in\mathbb{R}^{T\times F}$ and the clean wrapped phase spectrum $\bm{X}_p\in\mathbb{R}^{T\times F}$, respectively. The enhanced magnitude spectrum $\bm{\hat{X}}_m\in\mathbb{R}^{T\times F}$ is computed as:
\begin{align}
    \bm{\hat{X}}_m = ((\bm{Y}_m)^c \odot \bm{\hat{M}}^c)^{1/c},
\end{align}
where $\bm{\hat{M}}^c$ denotes the predicted clean compressed magnitude mask. 
The magnitude mask decoder comprises a dilated DenseNet, a 2D transposed convolution, an instance normalization, and a PReLU activation. This is followed by a deconvolution block reducing the amount of channels from $K$ to 1, and a learnable sigmoid function with $\beta=2$ \cite{fu2021metricgan+} estimating the magnitude mask. Similarly, the wrapped phase decoder consists of a dilated DenseNet, a 2D transposed convolution, an instance normalization, and a PReLU activation. This is followed by two parallel 2D convolutional layers predicting the pseudo-real and pseudo-imaginary part components. The clean wrapped phase spectrum is predicted using the two-argument arctangent function \cite{mp-senet}, yielding the enhanced wrapped phase spectrum $\bm{\hat{X}}_p$. The final enhanced waveform $\bm{\hat{x}}\in\mathbb{R}^D$ is recovered by applying an iSTFT to the enhanced magnitude spectrum $\bm{\hat{X}}_m$ and the enhanced wrapped phase spectrum $\bm{\hat{X}}_p$. 
\begin{table*}[h]
    \centering
    \caption{In-domain and out-of-domain speech enhancement performance. Models are trained and  \textit{VB-DemandEx}.}

\begin{adjustbox}{width=1\textwidth}
\begin{tabular}{@{}lc|cccc|cccc|cccc@{}}
    \toprule
        \multirow{2}{*}{Dataset} & & \multicolumn{4}{c|}{In-Domain} & \multicolumn{8}{c}{\: \: \: Out-Of-Domain} \\ \cmidrule(lr){3-6}\cmidrule(l){7-14}
        & & \multicolumn{4}{c|}{\textit{VB-DemandEx}} & \multicolumn{4}{c|}{\textit{DNS 2020} Without Reverb} & \multicolumn{4}{c}{\textit{EARS-WHAM\_v2}}\\ \hline
    Model &\makecell{Params} & PESQ$\uparrow$        & SSNR$\uparrow$                & ESTOI$\uparrow$       & SI-SDR$\uparrow$          &PESQ$\uparrow$         & SSNR$\uparrow$                & ESTOI$\uparrow$       & SI-SDR$\uparrow$ & PESQ$\uparrow$         & SSNR$\uparrow$                & ESTOI$\uparrow$       & SI-SDR$\uparrow$   \\ \hline
    Noisy & - & 1.63   & -1.07   & 0.63   & 4.98      & 1.58    & 6.22    & 0.81 & 9.07 & {1.24} & {-0.80} & {0.64} & {5.36}   \\ \hline
    xLSTM-SENet \cite{xlstm-senet} & \SI{2.20}{M} & $2.97\scriptscriptstyle\pm 0.05$   & $7.93\scriptscriptstyle\pm 0.13$             & $0.80\scriptscriptstyle\pm 0.01$   & $16.41\scriptscriptstyle\pm 0.32$       & $1.72\scriptscriptstyle\pm 0.37$    & $3.25\scriptscriptstyle\pm 1.33$ &  $0.69\scriptscriptstyle\pm 0.10$ & $3.41\scriptscriptstyle\pm 3.48$  & {$1.51\scriptscriptstyle\pm 0.15$} & {$0.45\scriptscriptstyle\pm 0.57$} & {$0.56\scriptscriptstyle\pm 0.05$} & {$1.40\scriptscriptstyle\pm 2.14$} \\
    LSTM-SENet \cite{xlstm-senet}  & \SI{2.34}{M} & $3.00\scriptscriptstyle\pm 0.03$ & $7.98\scriptscriptstyle\pm 0.21$      &  $0.80\scriptscriptstyle\pm 0.00$ &  $16.64\scriptscriptstyle\pm 0.12$ & $1.98\scriptscriptstyle\pm 0.45$ & $4.90\scriptscriptstyle\pm 1.66$ & $0.72\scriptscriptstyle\pm 0.12$ & $4.75\scriptscriptstyle\pm 3.35$  & {$1.57\scriptscriptstyle\pm 0.18$} & {$0.85\scriptscriptstyle\pm 0.77$} & {$0.57\scriptscriptstyle\pm 0.08$} & {$1.92\scriptscriptstyle\pm 2.89$} \\
    SEMamba \cite{semamba}& \SI{2.25}{M} &$3.00\scriptscriptstyle\pm 0.02$ & $7.59\scriptscriptstyle\pm 0.18$      & $0.80\scriptscriptstyle\pm 0.00$ &$16.59\scriptscriptstyle\pm 0.16$       & $2.28\scriptscriptstyle\pm 0.13$  & $5.84\scriptscriptstyle\pm 1.03$ & $0.82\scriptscriptstyle\pm 0.03$ & $9.30\scriptscriptstyle\pm 1.58$  &  {$1.63\scriptscriptstyle\pm 0.05$}  & {$0.92\scriptscriptstyle\pm 0.51$} & {$0.60\scriptscriptstyle\pm 0.03$} & {$2.81\scriptscriptstyle\pm 0.52$}  \\
    MP-SENet \cite{mp-senet}& \SI{2.05}{M}  & $2.94\scriptscriptstyle\pm 0.07$ & $7.64\scriptscriptstyle\pm 0.28$ & $0.79\scriptscriptstyle\pm 0.01$ & $16.20\scriptscriptstyle\pm 0.32$       & $2.67\scriptscriptstyle\pm 0.01$ & $7.37\scriptscriptstyle\pm 0.38$ & $0.88\scriptscriptstyle\pm 0.01$ & $13.67\scriptscriptstyle\pm 0.89$  & {$1.86\scriptscriptstyle\pm 0.10$} & {$2.11\scriptscriptstyle\pm 0.27$} & {$0.68\scriptscriptstyle\pm 0.03$} & {$6.09\scriptscriptstyle\pm 0.67$}  \\
    \hline
    MambAttention& \SI{2.33}{M} & $3.03\scriptscriptstyle\pm 0.01$ & $7.67\scriptscriptstyle\pm 0.41$      &  $0.80\scriptscriptstyle\pm 0.00$ &  $16.68\scriptscriptstyle\pm 0.10$  &  $2.92\scriptscriptstyle\pm 0.12$ & $8.13\scriptscriptstyle\pm 0.73$ & $0.91\scriptscriptstyle\pm 0.01$ & $15.17\scriptscriptstyle\pm 1.36$ & {$2.01\scriptscriptstyle\pm 0.05$} & {$2.51\scriptscriptstyle\pm 0.22$} & {$0.73\scriptscriptstyle\pm 0.02$} & {$7.35\scriptscriptstyle\pm 0.45$}  \\
    \bottomrule
\end{tabular}
\end{adjustbox}

    \label{tab:generalization}
\end{table*}
\begin{figure*}[!tbh]
\centering
\begin{subfigure}[b]{1\textwidth}
   \includegraphics[width=1\textwidth]{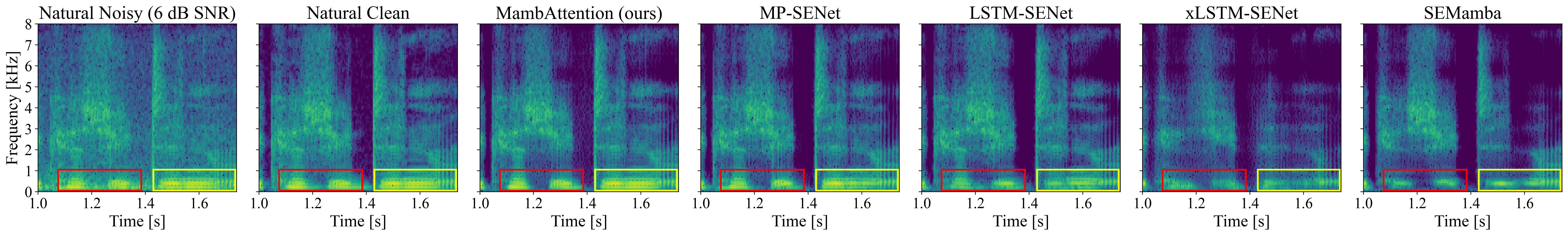}
   \caption{\textit{DNS 2020} without reverb (fileid\_90).}
   \label{fig:dns2020_spec} 
\end{subfigure}
\begin{subfigure}[b]{ 1\textwidth}
   \includegraphics[width=1\textwidth]{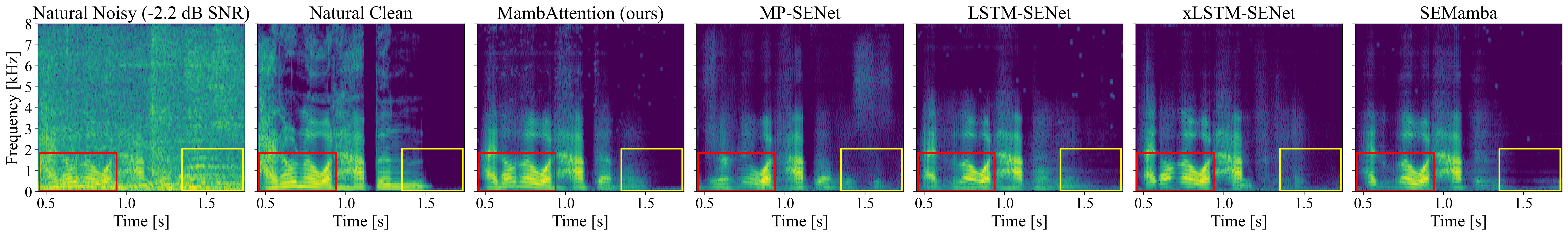}
   \caption{\textit{EARS-WHAM\_v2} (fileid\_00040).}
   \label{fig:ears_spec}
\end{subfigure}
\caption{Spectrogram visualizations of the noisy speech, clean speech, and enhanced speech from our proposed MambAttention and the Conformer, LSTM, xLSTM, and Mamba baselines.}
\label{fig:spectrograms}
\end{figure*}
\subsection{Loss Functions}
We follow \cite{mp-senet,semamba,xlstm-senet} and use a linear combination of loss functions. We use a time loss $\mathcal{L}_{\mathrm{Time}}$, magnitude loss $\mathcal{L}_{\mathrm{Mag.}}$, and complex loss $\mathcal{L}_{\mathrm{Com.}}$ defined by:
\begin{align}
    \mathcal{L}_{\mathrm{Time}} &= \mathbb{E}_{\bm{x},\bm{\hat{x}}}[\norm{\bm{x}-\bm{\hat{x}}}_1],\\
    \mathcal{L}_{\mathrm{Mag.}} &= \mathbb{E}_{\bm{X}_m,\bm{\hat{X}_m}}[\norm{{\bm{X}_m-\bm{\hat{X}_m}}}_2^2], \\
    \mathcal{L}_{\mathrm{Com.}} &=\mathbb{E}_{\bm{X}_r,\bm{\hat{X}_r}}[\norm{{\bm{X}_r-\bm{\hat{X}_r}}}_2^2] + \mathbb{E}_{\bm{X}_i,\bm{\hat{X}_i}}[\norm{{\bm{X}_i-\bm{\hat{X}_i}}}_2^2],
\end{align}
where $\mathbb{E}[\cdot]$ is the expectation operator, $\bm{x}\in\mathbb{R}^D$ is the clean speech, $\bm{\hat{x}}\in\mathbb{R}^D$ is the enhanced speech, and the pairs $(\bm{X}_r,\bm{{X}}_i)$ and $(\bm{\hat{X}}_r,\bm{\hat{X}}_i)$ are the real and imaginary parts of the clean complex spectrum $\bm{X}=\bm{X}_m\cdot \mathrm{e}^{j\bm{X}_p}\in\mathbb{C}^{T\times F}$ and the enhanced complex spectrum $\bm{\hat{X}}=\bm{X}_m\cdot \mathrm{e}^{j\bm{\hat{X}}_p}\in\mathbb{C}^{T\times F}$, respectively. Additionally, we use the instantaneous phase loss $\mathcal{L}_{\mathrm{IP}}$, group delay loss $\mathcal{L}_{\mathrm{GD}}$, and instantaneous angular frequency loss $\mathcal{L}_{\mathrm{IAF}}$ presented in \cite{ai2023neural} and define the phase loss as:
\begin{align}
    \mathcal{L}_{\mathrm{Pha.}} = \mathcal{L}_{\mathrm{IP}} + \mathcal{L}_{\mathrm{GD}} + \mathcal{L}_{\mathrm{IAF}}.
\end{align}
To improve training stability, we employ the consistency loss $\mathcal{L}_{\mathrm{Con.}}$ presented in \cite{zadorozhnyy23_interspeech}:
\begin{multline}
    \mathcal{L}_{\mathrm{Con.}} = \mathbb{E}_{\bm{\hat{X}}_r}[\norm{\bm{\hat{X}}_r - \mathrm{STFT}(\mathrm{iSTFT}(\bm{\hat{X}}_r))}_2^2] \\
    + \mathbb{E}_{\bm{\hat{X}}_i}[\norm{\bm{\hat{X}}_i - \mathrm{STFT}(\mathrm{iSTFT}(\bm{\hat{X}}_i))}_2^2].
\end{multline}
Finally, we use the metric discriminator $D$ from \cite{abdulatif2024cmgan} for adversarial training, which uses perceptual evaluation of speech quality (PESQ) as the target objective metric. The PESQ scores are linearly normalized to $[0,1]$. The discriminator loss $\mathcal{L}_{\mathrm{D}}$ and its corresponding PESQ-based generator loss $\mathcal{L}_{\mathrm{PESQ}}$ are given by:
\begin{multline}
    \mathcal{L}_{\mathrm{D}} = \mathbb{E}_{\bm{X}_m}[\norm{D(\bm{X}_m,\bm{X}_m) -1}_2^2] \\
    + \mathbb{E}_{\bm{X}_m,\bm{\hat{X}}_m}[\norm{D(\bm{X}_m,\bm{\hat{X}}_m)-Q_{\mathrm{PESQ}}}_2^2],
\end{multline}
and
\begin{align}
    \mathcal{L}_{\mathrm{PESQ}} = \mathbb{E}_{\bm{X}_m,\bm{\hat{X}}_m}[\norm{D(\bm{X}_m,\bm{\hat{X}}_m)-1}_2^2],
\end{align}
where $Q_{\mathrm{PESQ}}\in[0,1]$ is the scaled PESQ score between $\bm{X}_m$ and $\bm{\hat{X}}_m$. The final generator loss $\mathcal{L}_{\mathrm{G}}$ then becomes a linear combination of the above-mentioned loss functions:
\begin{multline}
    \mathcal{L}_{\mathrm{G}} = \alpha_1 \mathcal{L}_{\mathrm{Time}} + \alpha_2 \mathcal{L}_{\mathrm{Mag.}} + \alpha_3 \mathcal{L}_{\mathrm{Com.}} \\
    + \alpha_4 \mathcal{L}_{\mathrm{Pha.}} + \alpha_5\mathcal{L}_{\mathrm{Con.}} + \alpha_6 \mathcal{L}_{\mathrm{PESQ}}.
\end{multline}
During training, $\mathcal{L}_{\mathrm{G}}$ and $\mathcal{L}_{\mathrm{D}}$ are jointly minimized.
\subsection{Hyperparameter Settings}
All discriminative baselines are trained using the officially provided code. To reduce training time and vRAM usage, we follow \cite{mp-senet,semamba,xlstm-senet, yan2025lisennet} and train all models on randomly cropped \SI{2}{s} audio clips. Unless audio files are natively sampled at \SI{16}{kHz}, they are downsampled to \SI{16}{kHz}. For MambAttention, we use an FFT order of 400, a Hann window size of 400, and a hop size of 100 for all STFTs. Moreover, we use a magnitude spectrum compression factor of $c=0.3$ \cite{mp-senet}. We follow MP-SENet \cite{mp-senet}, SEMamba \cite{semamba}, and xLSTM-SENet \cite{xlstm-senet}, and fix the model feature dimension and number of channels $d_m=K=64$, the amount of layers $R=4$, and the expansion factor to $E_f=4$ for our MambAttention model. 
Like MP-SENet, we use $h=8$ attention heads in our proposed MambAttention model. We follow  \cite{lu2023explicit}, and set the hyperparameters in the generator loss function to $\alpha_1=0.2,\alpha_2=0.9,\alpha_3=0.1,\alpha_4=0.3,\alpha_5=0.1,$ and $\alpha_6=0.05$. All models trained on \textit{VB-DemandEx} and \textit{DNS 2020} are trained for \SI{550}{k} and \SI{950}{k} steps respectively, with a batch size of $M=8$ on four AMD MI250X GPUs, as validation performance stops improving when training longer. For both the generator and discriminator, we use AdamW \cite{Loshchilov2017DecoupledWD} with an initial learning rate of 0.0005, a weight decay of 0.01, $\beta_1=0.8$, and $\beta_2=0.99$. We use an exponential learning rate scheduler with a learning rate decay of $0.99$. For evaluation, we select the checkpoint with the highest PESQ score on the validation set, saved every 250 steps.

\subsection{Evaluation Metrics}

To assess the quality of the enhanced speech of discriminative models, we apply wide-band PESQ \cite{pesq}, which reports values between -0.5 (poor) and 4.5 (excellent). Additionally, we report the standard waveform-matching-based evaluation metrics SSNR \cite{hansen1998effective} and scale-invariant signal-to-distortion ratio (SI-SDR) \cite{le2019sdr}. Both SSNR and SI-SDR are reported in dB. To predict the intelligibility of the enhanced speech, we use extended short-time objective intelligibility (ESTOI) \cite{jensen2016algorithm}, which effectively reports values between 0 and 1.

It is well known that reference-signal based metrics such as PESQ, ESTOI, SI-SDR, and SSNR may be less suited for assessing the speech enhancement performance of generative models for example due to the lack of waveform-level alignment between the reference and enhanced speech signals \cite{wang2024selm, yang2024uniaudio}. Thus, to evalute the quality of the enhanced speech of generative models, we employ the reference-free perceptual quality estimator DNSMOS \cite{reddy2022dnsmos}, which outputs three scores between 1 and 5 for the signal quality (SIG), the background noise (BAK), and the overall quality (OVRL).

For all measures, higher values indicate better performance. All models and baselines trained by us are trained with 5 different seeds, and we report the mean and standard deviation.
\section{RESULTS}\label{sec:results}
\begin{table*}[h]
    \centering
    \caption{DNSMOS scores on the \textit{DNS 2020} test set with and without reverberation and the real-recordings test set. “-” denotes that the result is not provided in the original paper and could not be reproduced.}
    \begin{adjustbox}{width=1\textwidth}
\begin{tabular}{lccccccccccc}
\toprule
\multirow{3}{*}{Model} & \multirow{3}{*}{Type} & \multirow{3}{*}{Params} &
  \multicolumn{3}{c}{Without Reverb} & \multicolumn{3}{c}{Real Recordings} & \multicolumn{3}{c}{With Reverb} \\
 \cmidrule(lr){4-6}  \cmidrule(lr){7-9} \cmidrule(lr){10-12}
& & & SIG$\uparrow$& BAK$\uparrow$ & OVRL$\uparrow$ & SIG$\uparrow$& BAK$\uparrow$ & OVRL$\uparrow$ & SIG$\uparrow$ & BAK$\uparrow$ & OVRL$\uparrow$ \\
\midrule
Noisy & - & -  &3.39 & 2.62 & 2.48 & 3.05 & 2.51 & 2.26& 1.76 & 1.50 & 1.39 \\
\hline
CDiffuse \cite{lu2022conditional} & Diffusion &\SI{2.64}{M}&3.29 & 3.64 & 3.05 &3.20 & 3.10 & 2.78 & 2.54 & 2.30 & 2.19 \\
SGMSE \cite{SGMSE} & Diffusion &\SI{3.56}{M} & 3.50 & 3.71 & 3.14 &3.30 & 2.89 & 2.79& 2.73 & 2.74 & 2.43\\
StoRM \cite{lemercier2023storm} & Diffusion &\SI{27.8}{M}& 3.51 & 3.94 & 3.21 & 3.41 & 3.38 & 2.94 & 2.95 &3.14 & 2.52 \\
\hline
SELM \cite{wang2024selm} & LM &- & 3.51 & 4.10 & 3.26 & 3.59 & 3.44 & 3.12 & 3.16 & 3.58 & 2.70\\
MaskSR \cite{masksr}& LM &\SI{145}{M}& 3.59 & 4.12 & 3.34  & 3.43 & 4.03 & 3.14& 3.53 & 4.07 & 3.25 \\
AnyEnhance \cite{anyenhance}& LM &\SI{363.54}{M} & 3.64 & 4.18 & 3.42 & 3.49 & 3.98 & 3.16& 3.50 & 4.04 & 3.20\\
GenSE \cite{yao2025gense}& LM &- & 3.65 & 4.18 & 3.43 & - & - & - & 3.49 & 3.73 & 3.19\\
LLaSE-G1 \cite{llase}& LM &\SI{1.07}{B} & 3.66 & 4.17 & 3.42 & 3.41 & 3.91 & 3.08 &3.59 & 4.10 & 3.33\\
\hline
MambAttention (\textit{DNS 2020)} & Regression &\SI{2.33}{M} &$3.62\scriptstyle\pm 0.00$ & $4.19\scriptstyle\pm 0.00$ & $3.40\scriptstyle\pm 0.00$ & $3.46\scriptstyle\pm 0.00$ & $4.06\scriptstyle\pm 0.01$ & $3.17\scriptstyle\pm 0.00$& $2.88\scriptstyle\pm 0.02$ & $3.41\scriptstyle\pm 0.02$ & $2.37\scriptstyle\pm 0.01$\\
MambAttention (\textit{VB-DemandEx)} & Regression &\SI{2.33}{M} &$3.57\scriptstyle\pm 0.00$ & $4.16\scriptstyle\pm 0.00$ & $3.35\scriptstyle\pm 0.02$ & $3.25\scriptstyle\pm 0.01$ & $3.90\scriptstyle\pm 0.01$ & $2.90\scriptstyle\pm 0.01$& $2.66\scriptstyle\pm 0.02$ & $3.71\scriptstyle\pm 0.03$ & $2.16\scriptstyle\pm 0.03$\\
\bottomrule
\end{tabular}
\end{adjustbox}

    \label{tab:llm+diff}
\end{table*}
\begin{table*}[h]
    \centering
    \caption{Ablation study. Default configurations for our MambAttention model are the T- and F-MHA modules before the T- and F-Mamba blocks, respectively, shared weights between the T- and F-MHA modules, and processing across the time-dimension before the frequency-dimension. Models are trained on \textit{VB-DemandEx}.}

\begin{adjustbox}{width=1\textwidth}
\begin{tabular}{@{}lc|cccc|cccc|cccc@{}}
    \toprule
        \multirow{2}{*}{Dataset} & & \multicolumn{4}{c|}{In-Domain} & \multicolumn{8}{c}{\: \: \: Out-Of-Domain} \\ \cmidrule(lr){3-6}\cmidrule(l){7-14}
        & & \multicolumn{4}{c|}{\textit{VB-DemandEx}} & \multicolumn{4}{c|}{\textit{DNS 2020} Without Reverb} & \multicolumn{4}{c}{\textit{EARS-WHAM\_v2}}\\ \hline
    Model &\makecell{Params} & PESQ$\uparrow$         & SSNR$\uparrow$                & ESTOI$\uparrow$       & SI-SDR$\uparrow$          &PESQ$\uparrow$         & SSNR$\uparrow$                & ESTOI$\uparrow$       & SI-SDR$\uparrow$ & PESQ$\uparrow$         & SSNR$\uparrow$                & ESTOI$\uparrow$       & SI-SDR$\uparrow$   \\ \hline
     Noisy & - & 1.63   & -1.07   & 0.63   & 4.98      & 1.58    & 6.22    & 0.81 & 9.07 & {1.24} & {-0.80} & {0.64} & {5.36}   \\ \hline
MambAttention& \SI{2.33}{M} & $3.03\scriptscriptstyle\pm 0.01$ & $7.67\scriptscriptstyle\pm 0.41$      &  $0.80\scriptscriptstyle\pm 0.00$ &  $16.68\scriptscriptstyle\pm 0.10$  &  $2.92\scriptscriptstyle\pm 0.12$ & $8.13\scriptscriptstyle\pm 0.73$ & $0.91\scriptscriptstyle\pm 0.01$ & $15.17\scriptscriptstyle\pm 1.36$ & {$2.01\scriptscriptstyle\pm 0.05$} & {$2.51\scriptscriptstyle\pm 0.22$} & {$0.73\scriptscriptstyle\pm 0.02$} & {$7.35\scriptscriptstyle\pm 0.45$}  \\
     Attention after& \SI{2.33}{M} & $3.03\scriptscriptstyle\pm 0.01$ & $7.77\scriptscriptstyle\pm 0.16$      &  $0.80\scriptscriptstyle\pm 0.00$ &  $16.71\scriptscriptstyle\pm 0.06$  &  $2.71\scriptscriptstyle\pm 0.20$ & $7.23\scriptscriptstyle\pm 0.98$ & $0.87\scriptscriptstyle\pm 0.03$ & $12.86\scriptscriptstyle\pm 2.30$ & {$1.93\scriptscriptstyle\pm 0.14$} & {$1.70\scriptscriptstyle\pm 0.37$} & {$0.68\scriptscriptstyle\pm 0.05$} & {$4.84\scriptscriptstyle\pm 0.69$}  \\
     w/o weight sharing& \SI{2.39}{M} & $3.04\scriptscriptstyle\pm 0.01$ & $7.95\scriptscriptstyle\pm 0.07$      &  $0.80\scriptscriptstyle\pm 0.00$ &  $16.71\scriptscriptstyle\pm 0.10$  &  $2.61\scriptscriptstyle\pm 0.23$ & $6.76\scriptscriptstyle\pm 1.23$ & $0.86\scriptscriptstyle\pm 0.05$ & $11.58\scriptscriptstyle\pm 2.71$ & {$1.88\scriptscriptstyle\pm 0.07$} & {$1.61\scriptscriptstyle\pm 0.32$} & {$0.67\scriptscriptstyle\pm 0.02$} & {$4.45\scriptscriptstyle\pm 1.00$} \\
     w/o MHA modules \cite{semamba}& \SI{2.25}{M} &$3.00\scriptscriptstyle\pm 0.02$ & $7.59\scriptscriptstyle\pm 0.18$      & $0.80\scriptscriptstyle\pm 0.00$ &$16.59\scriptscriptstyle\pm 0.16$       & $2.28\scriptscriptstyle\pm 0.13$  & $5.84\scriptscriptstyle\pm 1.03$ & $0.82\scriptscriptstyle\pm 0.03$ & $9.30\scriptscriptstyle\pm 1.58$  &  {$1.63\scriptscriptstyle\pm 0.05$}  & {$0.92\scriptscriptstyle\pm 0.51$} & {$0.60\scriptscriptstyle\pm 0.03$} & {$2.81\scriptscriptstyle\pm 0.52$}  \\
     {Frequency $\to$ Time}& \SI{2.33}{M} & {$3.02\scriptscriptstyle\pm 0.03$} & {$7.65\scriptscriptstyle\pm 0.20$}      &  {$0.80\scriptscriptstyle\pm 0.01$} &  {$16.57\scriptscriptstyle\pm 0.17$}  &  {$2.55\scriptscriptstyle\pm 0.40$} & {$6.27\scriptscriptstyle\pm 1.40$} & {$0.84\scriptscriptstyle\pm 0.07$} & {$11.28\scriptscriptstyle\pm 4.25$} & {$1.85\scriptscriptstyle\pm 0.24$} & {$1.53\scriptscriptstyle\pm 0.80$} & {$0.65\scriptscriptstyle\pm 0.09$} & {$4.30\scriptscriptstyle\pm 3.37$} \\
    \bottomrule
\end{tabular}
\end{adjustbox}

    \label{tab:ablations}
\end{table*}
\subsection{Generalization Performance}\label{subsec:gen_perf}
In \autoref{tab:generalization}, we report in- and out-of-domain speech enhancement performance of our MambAttention model as well as the discriminative state-of-the-art LSTM-, xLSTM-, Mamba-, and Conformer-based baselines \cite{xlstm-senet, semamba, mp-senet}. For simplicity, we rename the LSTM baseline to
LSTM-SENet.

We observe that our proposed MambAttention model outperforms all other discriminative baselines on the out-of-domain datasets on all reported metrics. This indicates superior generalization performance compared to existing models, as both recording conditions, noise, and speaker types are significantly different across both out-of-domain datasets compared to our \textit{VB-DemandEx} dataset. Compared to the Mamba-based SEMamba \cite{semamba}, we note that adding the shared MHA modules greatly improves out-of-domain performance, while only adding approximately \SI{3.4}{\percent} additional parameters. For example, on \textit{DNS 2020} without reverb, PESQ increases by 0.64, SSNR increases by 2.29, ESTOI increases by 0.09, and SI-SDR increases by 5.87 from using our MambAttention model over the pure Mamba baseline. Moreover, we observe that the recurrent sequence models LSTM and xLSTM exhibit the worst generalization performance, which could indicate overfitting to the training set, or domain-specific information being accumulated in the hidden state as hypothesized in \cite{kim22f_interspeech,long2024dgmamba}. In fact, both LSTM-SENet and xLSTM-SENet yield worse SSNR, ESTOI, and SI-SDR scores on the out-of-domain \textit{DNS 2020} test set without reverb than the unprocessed noisy samples from the test set itself. SEMamba, while still utilizing a sequence model, shows substantially better generalization performance compared to LSTM-SENet and xLSTM-SENet; however, only the Conformer-based MP-SENet and our MambAttention model consistently improve all reported metrics on both out-of-domain datasets. Additionally, we observe that all models exhibit comparable performance on the in-domain \textit{VB-DemandEx} dataset. This aligns with previous findings on the \textit{VoiceBank+Demand} dataset, which features higher SNRs compared to \textit{VB-DemandEx} \cite{xlstm-senet}.

To support the claim of superior generalization performance of our MambAttention model, we visualize spectrograms of clean speech, noisy speech, and the speech waveforms enhanced by our MambAttention model and the discriminative LSTM, xLSTM, Mamba, and Conformer baselines on the out-of-domain \textit{DNS 2020} without reverb and \textit{EARS-WHAM\_v2} test sets. For each model in \autoref{fig:spectrograms}, we select the seed with the median PESQ score on each test set to provide a fair comparison. As seen in the yellow and red boxes in \autoref{fig:dns2020_spec}, our MambAttention model and the Conformer-based model are the only models that mostly reconstruct the fundamental harmonics. The red boxes in \autoref{fig:ears_spec} reveal the same effect at a significantly lower SNR, but to a larger extent. Interestingly, the yellow boxes in \autoref{fig:ears_spec} show that only our MambAttention model, and to some extent, the xLSTM- and Mamba-based baselines, almost reconstruct silence, whereas the Conformer- and LSTM-based baselines do not.

\begin{table*}[h]
    \centering
    \caption{In-domain and out-of-domain speech enhancement performance of LSTM- and xLSTM-based models with added MHA modules. Models are trained on \textit{VB-DemandEx}.}
    \begin{adjustbox}{width=1\textwidth}
\begin{tabular}{@{}lc|cccc|cccc|cccc@{}}
    \toprule
        \multirow{2}{*}{Dataset} & & \multicolumn{4}{c|}{In-Domain} & \multicolumn{8}{c}{\: \: \: Out-Of-Domain} \\ \cmidrule(lr){3-6}\cmidrule(l){7-14}
        & & \multicolumn{4}{c|}{\textit{VB-DemandEx}} & \multicolumn{4}{c|}{\textit{DNS 2020} Without Reverb} & \multicolumn{4}{c}{\textit{EARS-WHAM\_v2}}\\ \hline
    Model &\makecell{Params} & PESQ$\uparrow$         & SSNR$\uparrow$                & ESTOI$\uparrow$       & SI-SDR$\uparrow$          &PESQ$\uparrow$         & SSNR$\uparrow$                & ESTOI$\uparrow$       & SI-SDR$\uparrow$ & PESQ$\uparrow$         & SSNR$\uparrow$                & ESTOI$\uparrow$       & SI-SDR$\uparrow$   \\ \hline
    Noisy & - & 1.63   & -1.07   & 0.63   & 4.98      & 1.58    & 6.22    & 0.81 & 9.07 & {1.24} & {-0.80} & {0.64} & {5.36}   \\ \hline
        xLSTM-Attention & \SI{2.27}{M} & $3.02\scriptscriptstyle\pm 0.01$   & $7.69\scriptscriptstyle\pm 0.19$             & $0.80\scriptscriptstyle\pm 0.00$   & $16.65\scriptscriptstyle\pm 0.11$       & $2.80\scriptscriptstyle\pm 0.17$    & $7.19\scriptscriptstyle\pm 0.93$ &  $0.89\scriptscriptstyle\pm 0.03$ & $13.91\scriptscriptstyle\pm 1.89$  & {$1.93\scriptscriptstyle\pm 0.10$} & {$1.96\scriptscriptstyle\pm 0.38$} & {$0.68\scriptscriptstyle\pm 0.02$} & {$6.14\scriptscriptstyle\pm 1.28$} \\
    LSTM-Attention & \SI{2.48}{M} & $3.02\scriptscriptstyle\pm 0.04$ & $7.65\scriptscriptstyle\pm 0.34$      &  $0.80\scriptscriptstyle\pm 0.01$ &  $16.60\scriptscriptstyle\pm 0.28$ & $2.55\scriptscriptstyle\pm 0.18$ & $5.79\scriptscriptstyle\pm 0.88$ & $0.85\scriptscriptstyle\pm 0.03$ & $10.96\scriptscriptstyle\pm 1.62$  & {$1.86\scriptscriptstyle\pm 0.16$} & {$1.17\scriptscriptstyle\pm 0.49$} & {$0.66\scriptscriptstyle\pm 0.06$} & {$4.51\scriptscriptstyle\pm 1.81$} \\
    MP-SENet \cite{mp-senet}& \SI{2.05}{M}  & $2.94\scriptscriptstyle\pm 0.07$ & $7.64\scriptscriptstyle\pm 0.28$ & $0.79\scriptscriptstyle\pm 0.01$ & $16.20\scriptscriptstyle\pm 0.32$       & $2.67\scriptscriptstyle\pm 0.01$ & $7.37\scriptscriptstyle\pm 0.38$ & $0.88\scriptscriptstyle\pm 0.01$ & $13.67\scriptscriptstyle\pm 0.89$  & {$1.86\scriptscriptstyle\pm 0.10$} & {$2.11\scriptscriptstyle\pm 0.27$} & {$0.68\scriptscriptstyle\pm 0.03$} & {$6.09\scriptscriptstyle\pm 0.67$}  \\
    \hline
    MambAttention& \SI{2.33}{M} & $3.03\scriptscriptstyle\pm 0.01$ & $7.67\scriptscriptstyle\pm 0.41$      &  $0.80\scriptscriptstyle\pm 0.00$ &  $16.68\scriptscriptstyle\pm 0.10$  &  $2.92\scriptscriptstyle\pm 0.12$ & $8.13\scriptscriptstyle\pm 0.73$ & $0.91\scriptscriptstyle\pm 0.01$ & $15.17\scriptscriptstyle\pm 1.36$ & {$2.01\scriptscriptstyle\pm 0.05$} & {$2.51\scriptscriptstyle\pm 0.22$} & {$0.73\scriptscriptstyle\pm 0.02$} & {$7.35\scriptscriptstyle\pm 0.45$}  \\
    \bottomrule
\end{tabular}
\end{adjustbox}

    \label{tab:with_attn}
\end{table*}
In \autoref{tab:llm+diff}, we compare our proposed MambAttention model trained on \textit{DNS 2020} and on \textit{VB-DemandEx}, respectively, to generative diffusion model \cite{lu2022conditional, SGMSE, lemercier2023storm} and LM baselines \cite{wang2024selm, masksr, anyenhance, yao2025gense, llase}.
All diffusion model and speech LM baselines are trained on several thousand hours of data, and for at least two tasks (denoising and dereverberation), whereas MambAttention is only trained for denoising. Results for CDiffuse, SGMSE, StoRM, and SELM are taken from \cite{wang2024selm}. All other results are taken from the respective papers, or are replicated using the official provided code if possible.

\autoref{tab:llm+diff} demonstrates that MambAttention (\textit{DNS 2020}) substantially outperforms all diffusion model baselines on the \textit{DNS 2020} test set without reverb and the real-recordings test set. It is also competitive with the LM baselines on the \textit{DNS 2020} test set without reverb and reaches the highest BAK and OVRL across all models on the real recordings test set. Moreover, despite not being trained for dereverberation, MambAttention (\textit{DNS 2020}) achieves comparable performance to SGMSE on SIG and OVRL, while outperforming all diffusion model baselines on BAK. For MambAttention (\textit{VB-DemandEx}), all test sets in \autoref{tab:llm+diff} are out-of-domain and it is trained on a relatively small dataset ($<$\SI{10}{h}), yet it remains comparable to all LM baselines and better than all diffusion model baselines on the \textit{DNS 2020} test set without reverb. On the real-recordings test set, MambAttention (\textit{VB-DemandEx}) matches SGMSE on SIG and OVRL and surpasses all diffusion models on BAK. On the \textit{DNS 2020} test set with reverb, MambAttention (\textit{VB-DemandEx}) performs similar to CDiffuse on SIG and OVRL and outperforms all diffusion models and SELM while matching GenSE on BAK, despite not being trained for dereverberation. In summary, \autoref{tab:llm+diff}, demonstrates that our MambAttention model mostly outperforms diffusion model baselines and is competitive with large LM baselines, despite having significantly fewer parameters.

\begin{figure}[!t]
\centering
\begin{subfigure}[b]{0.50\textwidth}
   \includegraphics[width=1\columnwidth]{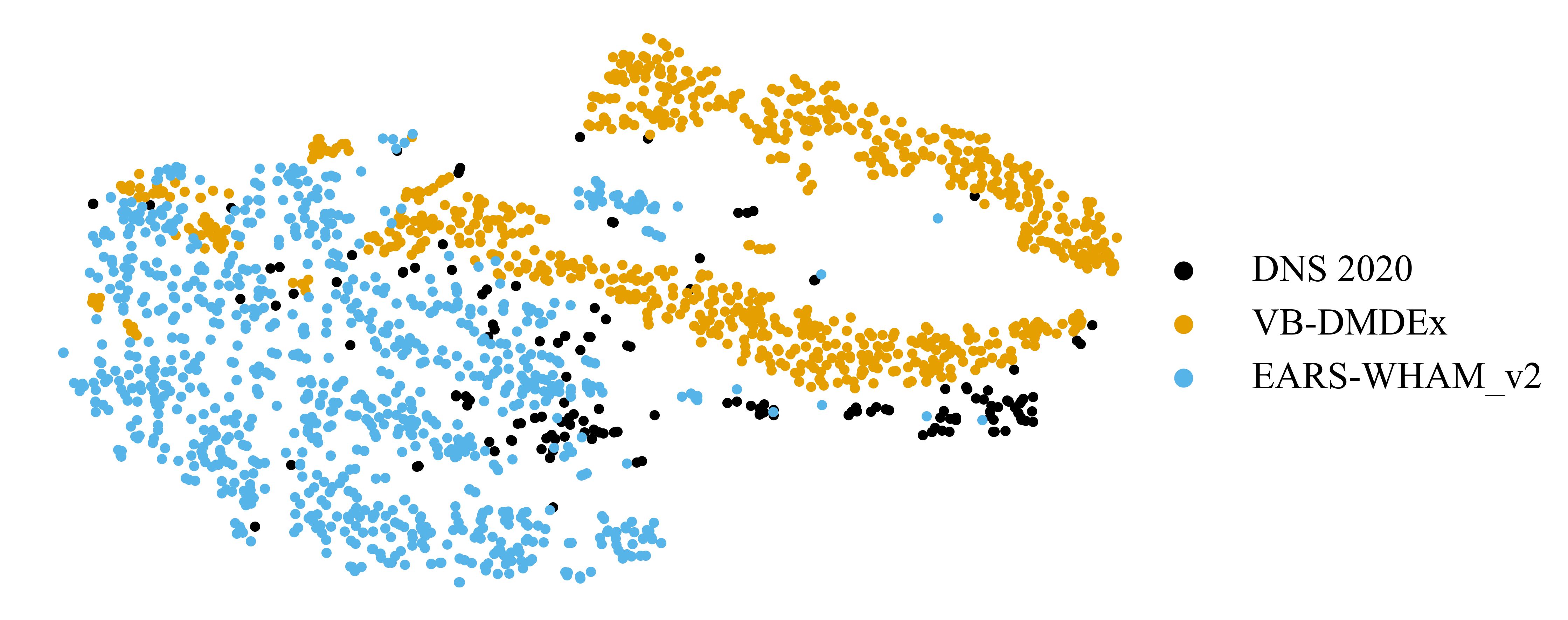}
   \caption{LSTM-SENet (LSTM).}
   \label{fig:lstm_tsne} 
\end{subfigure}

\begin{subfigure}[b]{0.5\textwidth}
   \includegraphics[width=1\columnwidth]{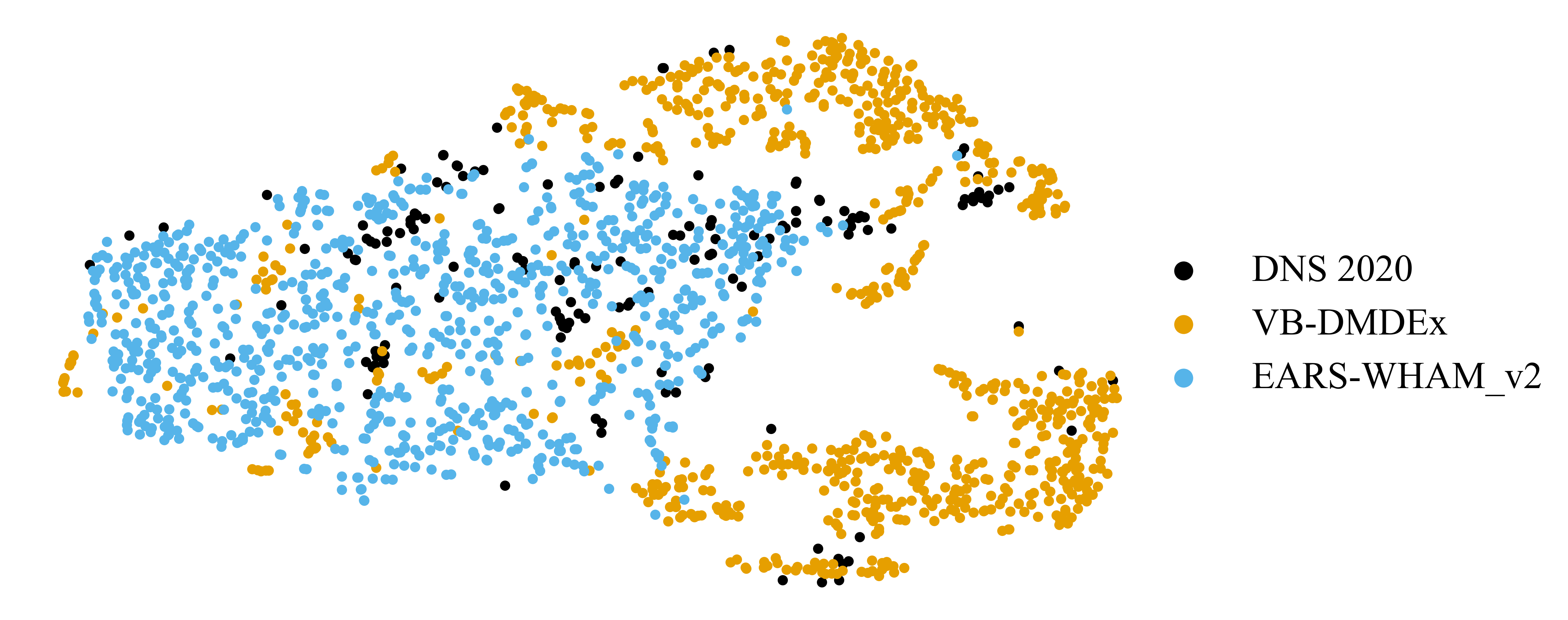}
   \caption{xLSTM-SENet (xLSTM).}
   \label{fig:xlstm_tsne}
\end{subfigure}

\begin{subfigure}[b]{0.5\textwidth}
   \includegraphics[width=1\columnwidth]{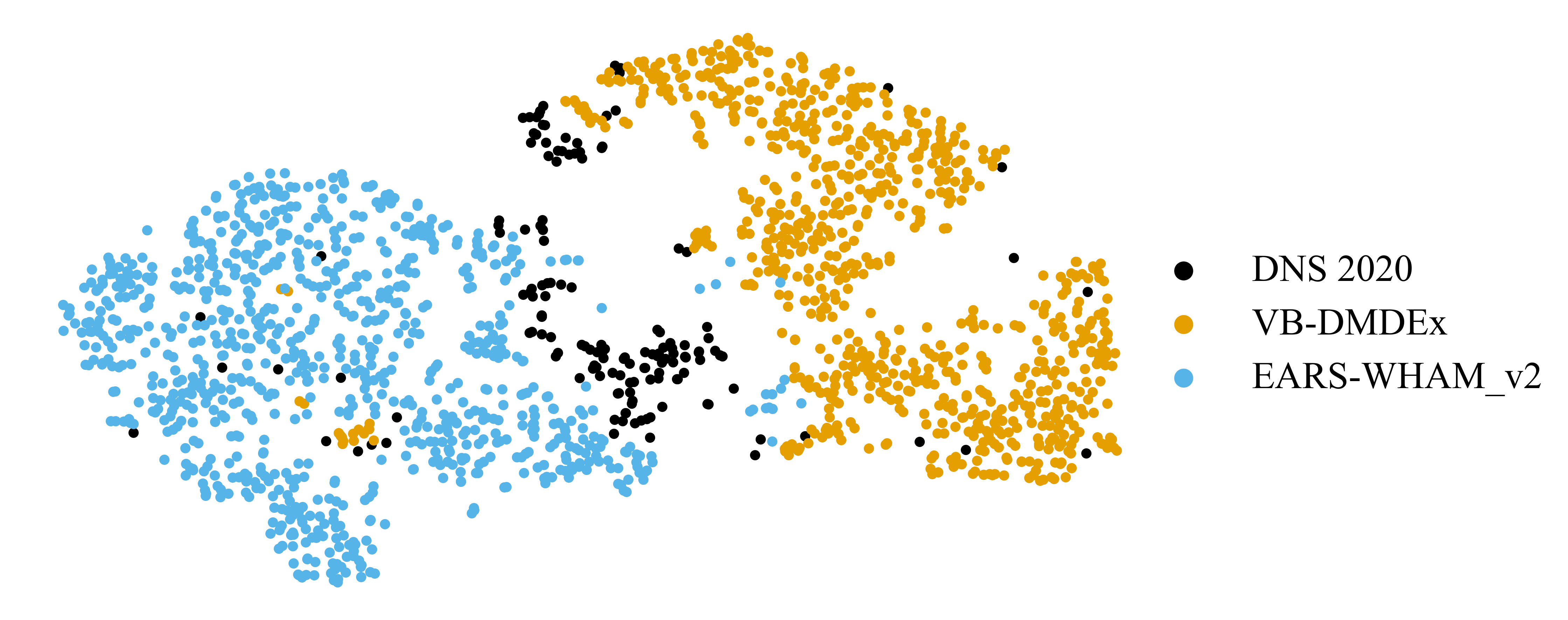}
   \caption{SEMamba (Mamba).}
   \label{fig:mamba_tsne}
\end{subfigure}

\begin{subfigure}[b]{0.5\textwidth}
   \includegraphics[width=1\columnwidth]{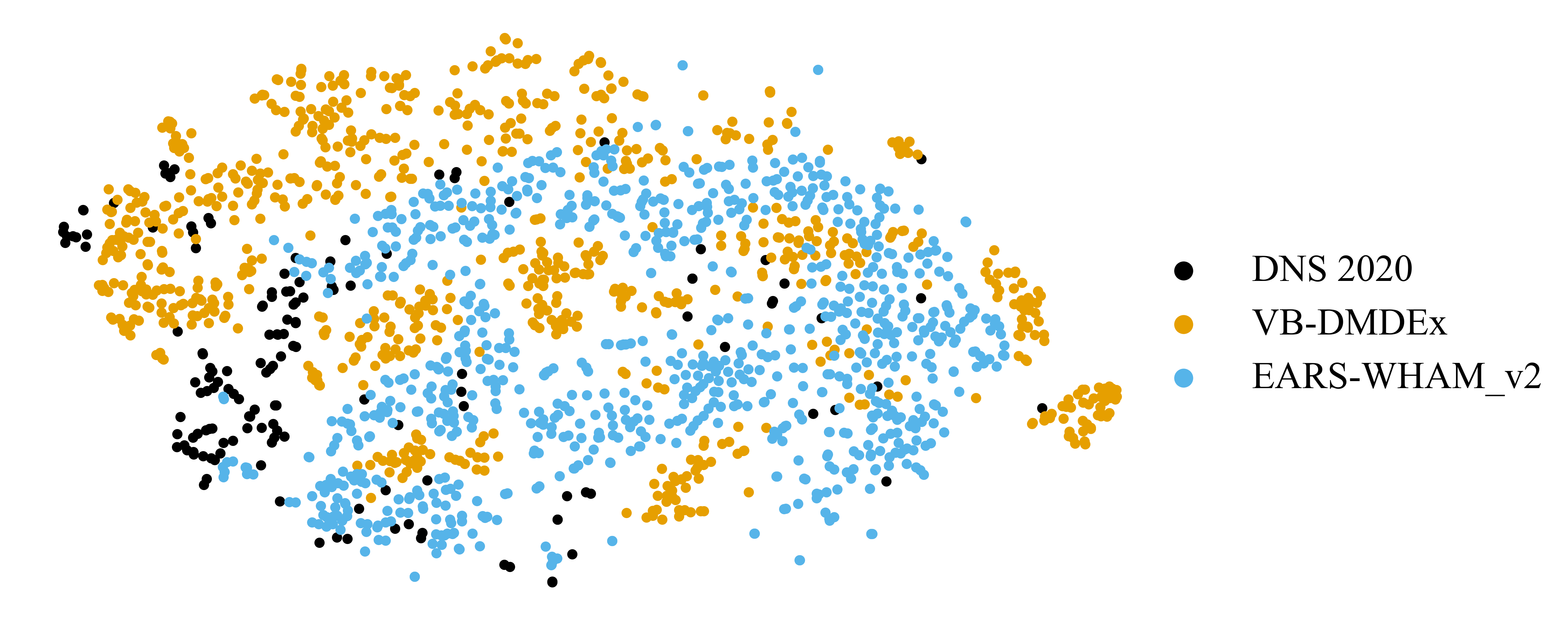}
   \caption{MP-SENet (Conformer).}
   \label{fig:conformer_tsne}
\end{subfigure}

\begin{subfigure}[b]{0.5\textwidth}
   \includegraphics[width=1\columnwidth]{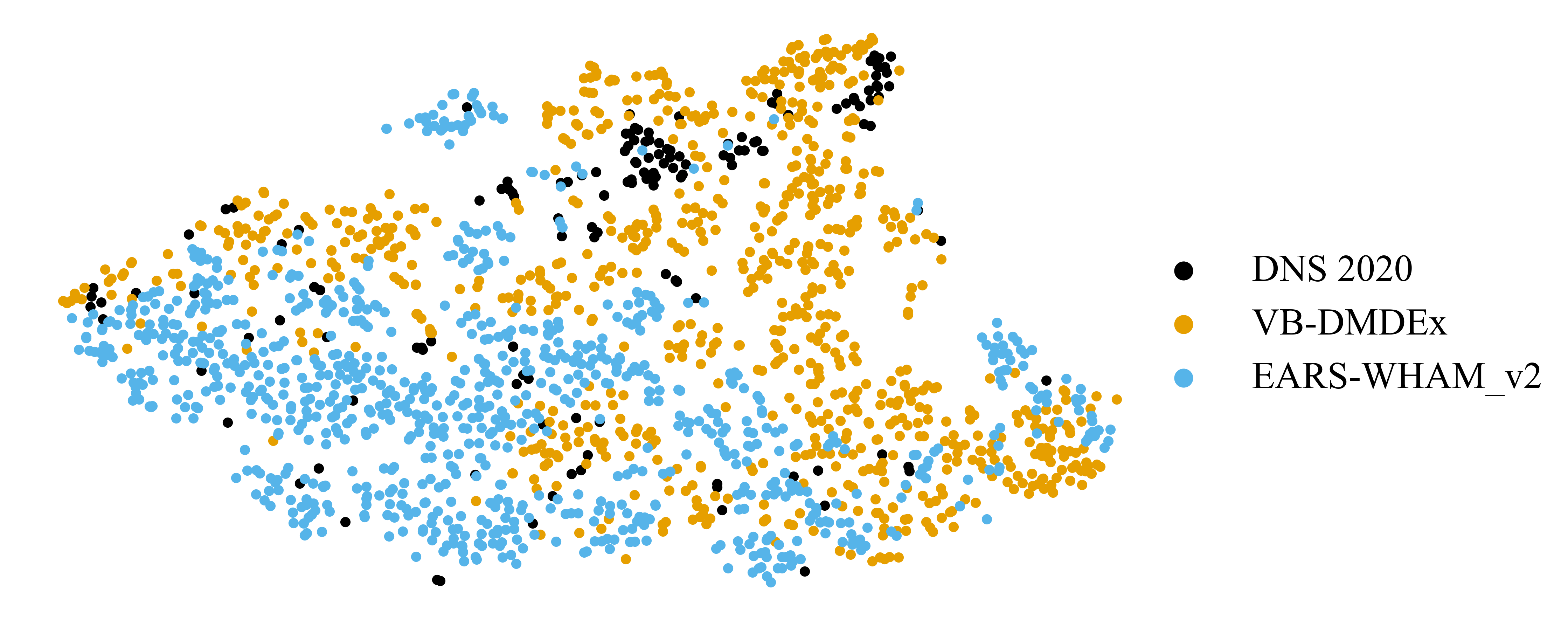}
   \caption{MambAttention (ours).}
   \label{fig:mamba_attention_tsne}
\end{subfigure}
\caption{t-SNE visualizations of the \textit{VB-DemandEx}, \textit{DNS 2020} without reverb, and \textit{EARS-WHAM\_v2} test sets.}
\label{fig:tsne_plots}
\end{figure}
\subsection{Ablation Study}
To investigate the performance impact of key architectural design choices, we conduct ablation studies on our proposed MambAttention block. 
As shown in \autoref{tab:ablations}, reversing the order of the MHA and Mamba blocks, negatively affects generalization performance on both out-of-domain datasets, as all reported performance metrics drop on both out-of-domain test sets. Hence, the ordering of components in our proposed MambAttention block affects the model's ability to generalize beyond the training distribution. 
Moreover, assigning separate weights to each T- and F-MHA module slightly increases parameter count but degrades generalization performance. This highlights the importance of the weight sharing mechanism for maintaining robustness to unseen speakers, noise types, and recording conditions. We hypothesize that weight sharing across the T- and F-MHA modules acts as a form of regularization. Instead of each individual MHA module attending to either time or frequency contents, we believe that the shared MHA modules force each layer of our MambAttention model to attend to time and frequency relations simultaneously. Thus, rather than overfitting to dataset-specific features, we believe the shared MHA modules encourage a focus on learning time and frequency structures that are more likely to generalize across various noise and speaker types. Finally, we flip the order of the time and frequency processing components in our MambAttention model. \autoref{tab:ablations} (Frequency $\to$ Time) demonstrates that processing along the frequency-dimension before the time-dimension yields significantly worse generalization performance, as all metrics on both out-of-domain test sets decrease significantly. These results align with CMGAN \cite{abdulatif2024cmgan}, which showed that processing across frequency before time in TS-Conformers yields worse speech enhancement performance.
We remark that although the ablation studies on the MHA modules in \autoref{tab:ablations} lead to reduced generalization performance, all variants still outperform the pure Mamba baseline from \cite{semamba}. 

We also integrate our T- and F-MHA modules into the xLSTM- and LSTM-based speech enhancement models from \cite{xlstm-senet} and denote them xLSTM-Attention and LSTM-Attention, respectively. For the xLSTM-Attention model, we replace the T and F-Mamba blocks in \autoref{fig: TF Mamba-Attention} with the T- and F-xLSTM blocks from \cite{xlstm-senet}. Similarly, for the LSTM-Attention model, we replace the T- and F-Mamba blocks in \autoref{fig: TF Mamba-Attention} with the T- and F-LSTM blocks from \cite{xlstm-senet}. Additionally, for the LSTM-Attention model, we reverse the order of the T-MHA and T-LSTM blocks and the order of the F-MHA and F-LSTM blocks, respectively. We choose these configurations for the xLSTM-Attention and LSTM-Attention models, as we found them to yield the best generalization performance. We remark that in \cite{pandey2022self}, the attention block was also placed after the LSTM block. In all cases, in-domain performance remains unchanged. As shown in \autoref{tab:with_attn}, both LSTM-Attention and xLSTM-Attention significantly outperform their MHA-free counterparts from \autoref{tab:generalization} on the two out-of-domain test sets. The xLSTM-Attention model consistently surpasses the Conformer baseline on the \textit{DNS 2020} test set without reverb and the \textit{EARS-WHAM\_v2} test set across all evaluation metrics except SSNR. Despite LSTM-Attention significanly surpassing its MHA-free baseline counterpart, the Conformer baseline remains superior on both out-of-domain test sets across all evaluation metrics used. These results are in line with \cite{pandey2022self}, where self-attending RNNs also were shown to improve cross-corpus generalization performance over their attention-free counterparts. We do, however, observe a significantly larger increase in generalization performance by adding our shared T- and F-MHA modules to the LSTM- and xLSTM-based models compared to \cite{pandey2022self}, which only uses a single self-attention block. Nevertheless, our proposed MambAttention model achieves superior generalization performance compared to both LSTM-Attention and xLSTM-Attention.

\subsection{Inspection of Latent Features}
In \autoref{tab:generalization}, we observe that the Conformer-based MP-SENet and our MambAttention model demonstrate significantly better generalization performance compared to the purely recurrent LSTM-, xLSTM-, and Mamba-based baselines. This is also consistently reflected in \autoref{tab:with_attn} when comparing attention-augmented baseline models with their attention-free counterparts from \autoref{tab:generalization}. To further understand the impact MHA has on generalization performance, we visually inspect the latent features produced by the LSTM-, xLSTM-, Mamba-, and Conformer-based models as well as our MambAttention model. Using t-Distributed Stochastic
Neighbor Embedding (t-SNE) \cite{cieslak2020tsne}, we visualize the outputs of the final LSTM, xLSTM, Mamba, Conformer, and MambAttention blocks, before they are passed to the magnitude mask and wrapped phase decoders. As we train all models with 5 different seeds, we select the seed with the median PESQ score on the out-of-domain \textit{DNS 2020} test set without reverb to provide a fair comparison. The t-SNE visualizations are done on the \textit{VB-DemandEx}, \textit{DNS 2020}, and \textit{EARS-WHAM\_v2} test sets.

As shown in \autoref{fig:tsne_plots}, the t-SNE embeddings for the in-domain and out-of-domain samples appear less tightly clustered and more intermingled across domains for the Conformer-based MP-SENet (\autoref{fig:conformer_tsne}) and our MambAttention model (\autoref{fig:mamba_attention_tsne}), compared to models with poorer generalization performance: LSTM-SENet (\autoref{fig:lstm_tsne}), xLSTM-SENet (\autoref{fig:xlstm_tsne}), and SEMamba (\autoref{fig:mamba_tsne}). For the LSTM, xLSTM, and Mamba models, the t-SNE embeddings of the individual test sets are significantly more clustered. At first sight, \autoref{fig:tsne_plots} may be surprising, however, the features in \autoref{fig:conformer_tsne} and \autoref{fig:mamba_attention_tsne} show that after being processed by the Conformer or our MambAttention blocks, t-SNE embeddings of both the in-domain and out-of-domain noisy speech are very close. This indicates that the learned features are less dataset-dependent, which supports our claim of superior generalization performance. Furthermore, this suggests that MHA may encourage the model to learn more dataset-invariant representations, rather than overfitting to dataset-specific patterns.

\begin{figure*}[!t]
\centering
\begin{subfigure}[h]{0.497\textwidth}
   \includegraphics[width=1\columnwidth]{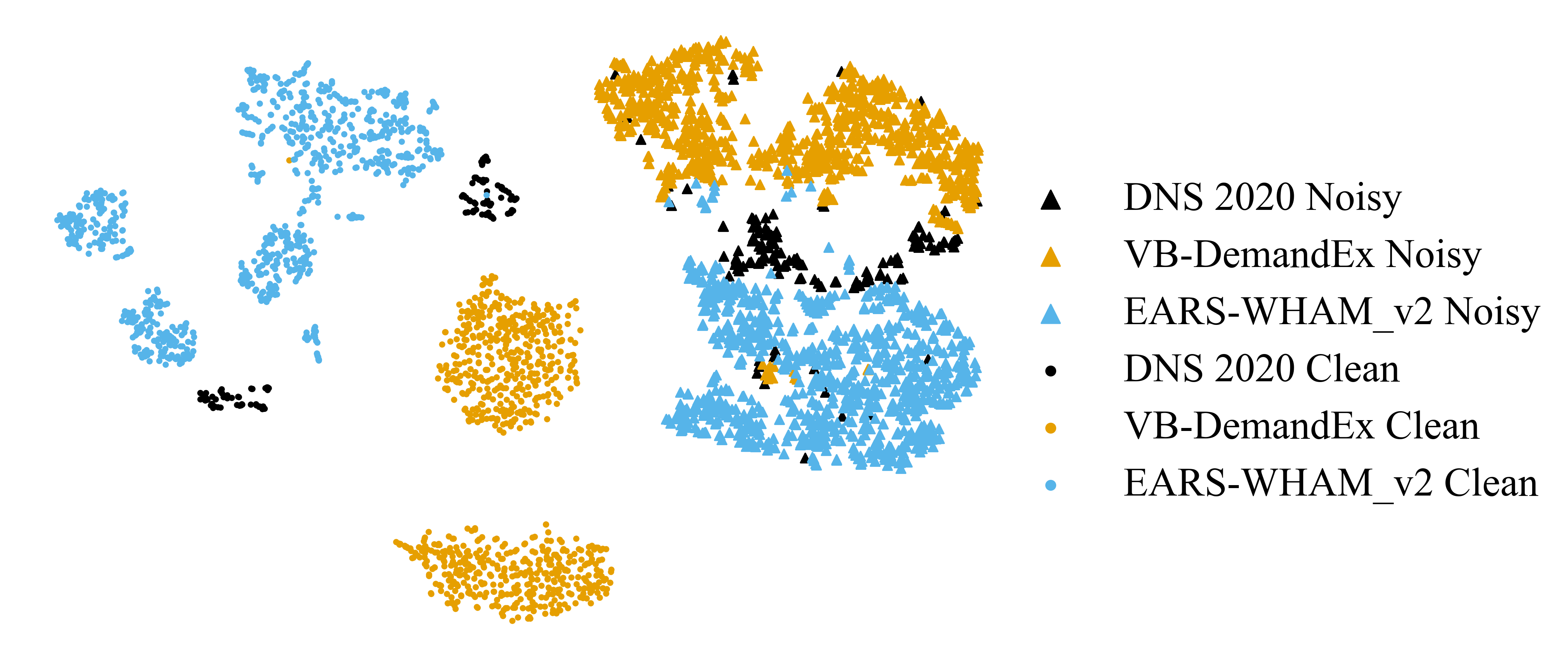}
   \caption{SEMamba (Mamba).}
   \label{fig:mamba_tsne+clean}
\end{subfigure}
\hfill
\begin{subfigure}[h]{0.497\textwidth}
   \includegraphics[width=1\columnwidth]{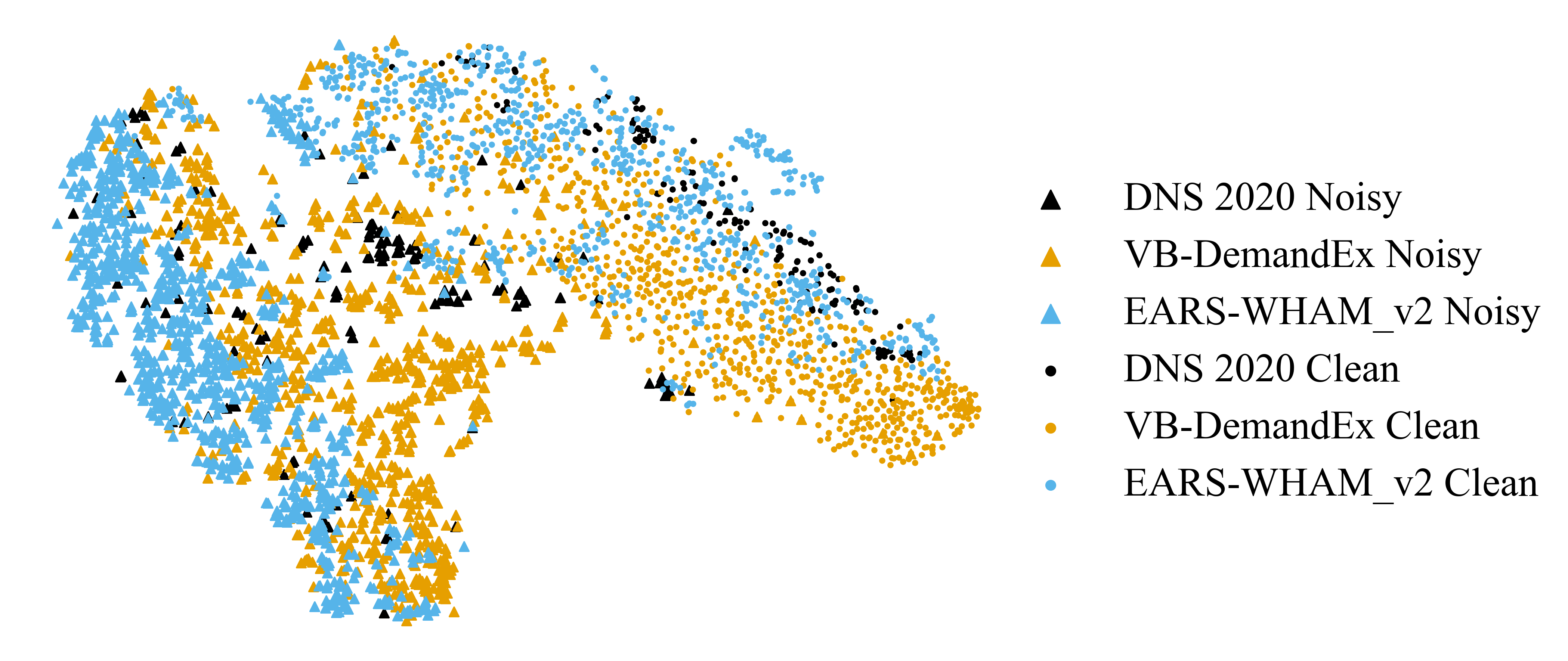}
   \caption{MambAttention (ours).}
   \label{fig:mamba_attention_tsne+clean}
\end{subfigure}
\caption{t-SNE visualizations of the \textit{VB-DemandEx}, \textit{DNS 2020} without reverb, and \textit{EARS-WHAM\_v2} test sets along with their clean references.}
\label{fig:tsne_plots+clean}
\end{figure*}

To gain further insights into the effect of the shared MHA modules in our MambAttention model compared to the pure Mamba baseline, we visualize t-SNE embeddings of the in- and out-of-domain \textit{VB-DemandEx}, \textit{DNS 2020} without reverb, and \textit{EARS-WHAM\_v2} test sets along with their clean references in \autoref{fig:tsne_plots+clean}. In \autoref{fig:mamba_attention_tsne+clean} we observe that after being processed by the MambAttention blocks, the t-SNE embeddings of the in- and out-of-domain clean references are clustered together. Moreover, the noisy speech is very close to their clean references in the t-SNE embedding space, suggesting that processed noisy speech and processed clean speech is similar. This indicates that the denoising process of the MambAttention blocks is effective, supporting the results presented in \autoref{tab:generalization}. This is in stark contrast to the pure Mamba model in \autoref{fig:mamba_tsne+clean}, where we observe that the t-SNE embeddings of both noisy speech and clean references are clearly separated and far apart.

\subsection{In-Domain Enhancement Results on DNS 2020}
As shown in \autoref{tab:generalization}, a range of neural architectures, including our proposed MambAttention model, achieve similar in-domain speech enhancement performance on our \textit{VB-DemandEx} dataset. Hence, in order to examine whether increased volume and data diversity better differentiate in-domain model performance, we train on the large-scale \textit{DNS 2020} dataset \cite{reddy20_interspeech}.
\begin{table}[H]
    \centering
    \caption{Speech Enhancement performance on the \textit{DNS 2020} test set without reverb.}
    \begin{adjustbox}{width=1\columnwidth}
\begin{tabular}{@{}lc|cccc@{}} 
\toprule
Model & \makecell{Params}& PESQ$\uparrow$ & SSNR$\uparrow$ & ESTOI$\uparrow$ & SI-SDR$\uparrow$  \\
\midrule
Noisy & - & $1.58$ & $6.22$ & $0.81$ & $9.07$ \\
\midrule
xLSTM-SENet \cite{xlstm-senet} & \SI{2.20}{M} & $3.59\scriptstyle\pm 0.02$ & $14.53\scriptstyle\pm 0.48$ & $0.95\scriptstyle\pm 0.00$ & $20.85\scriptstyle\pm 0.23$ \\
LSTM-SENet \cite{xlstm-senet} & \SI{2.34}{M} & $3.60\scriptstyle\pm 0.03$ & $15.02\scriptstyle\pm 0.17$ & $0.96\scriptstyle\pm 0.00$ & $21.00\scriptstyle\pm 0.22$ \\
SEMamba \cite{semamba} & \SI{2.25}{M} & $3.59\scriptstyle\pm 0.01$ & $14.83\scriptstyle\pm 0.47$ & $0.96\scriptstyle\pm 0.00$ & $21.04\scriptstyle\pm 0.12$ \\
MP-SENet \cite{mp-senet} & \SI{2.05}{M} & $3.61\scriptstyle\pm 0.02$ & $14.97\scriptstyle\pm 0.04$ & $0.95\scriptstyle\pm 0.00$ & $20.92\scriptstyle\pm0.02$  \\
\midrule
MambAttention & \SI{2.33}{M} & $3.67\scriptstyle\pm0.01$ & $15.12\scriptstyle\pm0.05$ & $0.96\scriptstyle\pm0.00$ & $21.23\scriptstyle\pm0.03$ \\
\bottomrule
\end{tabular}
\end{adjustbox}
    \label{tab:dns2020}
\end{table}
\begin{table*}[h]
    \centering
    \caption{Training time on the \textit{VB-DemandEx} and \textit{DNS 2020} training sets measured in GPU hours (wall clock time $\times$ number of GPUs) and inference time (wall clock time) measured in seconds on the \textit{VB-DemandEx}, \textit{DNS 2020} without reverb and \textit{EARS-WHAM\_v2} test sets.}
    \begin{adjustbox}{width=1\textwidth}
\begin{tabular}{@{}lcc|cc|ccc@{}}
    \toprule
    \multirow{2}{*}{Model} & \multirow{2}{*}{\makecell{Params}} & \multirow{2}{*}{\makecell{FLOPs}$\downarrow$} 
          & \multicolumn{2}{c|}{Training Time (per dataset)$\downarrow$} 
          & \multicolumn{3}{c}{Inference Time (per dataset)$\downarrow$} \\ 
    \cmidrule(lr){4-5} \cmidrule(lr){6-8}
          &  &  & \textit{VB-DemandEx} & \textit{DNS 2020} Without Reverb 
              & \textit{VB-DemandEx} & \textit{DNS 2020} Without Reverb & \textit{EARS-WHAM\_v2} \\ 
    \hline
    xLSTM-SENet \cite{xlstm-senet} & \SI{2.20}{M}  & \SI{80.71}{G} &  862 GPU Hours & 1499 GPU Hours & \SI{142}{s} & \SI{215}{s} & \SI{2580}{s} \\
    LSTM-SENet \cite{xlstm-senet} & \SI{2.34}{M} & \SI{88.59}{G}  & 131 GPU Hours & 225 GPU Hours & \SI{28}{s} & \SI{37}{s} & \SI{214}{s} \\
    SEMamba \cite{semamba} & \SI{2.25}{M}  & \SI{64.46}{G} & 187 GPU Hours & 325 GPU Hours & \SI{36}{s} & \SI{48}{s} & \SI{244}{s} \\
    MP-SENet \cite{mp-senet} & \SI{2.05}{M} & \SI{74.29}{G} & 388 GPU Hours & 675 GPU Hours & \SI{58}{s} & \SI{96}{s} & \SI{790}{s} \\
    \hline
    MambAttention & \SI{2.33}{M} & \SI{65.52}{G} & 212 GPU Hours & 365 GPU Hours & \SI{43}{s} & \SI{68}{s} & \SI{697}{s} \\
    \bottomrule
\end{tabular}
\end{adjustbox}

    \label{tab:runtime}
\end{table*}
\autoref{tab:dns2020} shows in-domain speech enhancement performance on the \textit{DNS 2020} dataset without reverb. We observe that when trained on \textit{DNS 2020}, compared to the pure Mamba baseline, our MambAttention model yields a PESQ score which is bigger by 0.08 and SI-SDR is bigger by 0.19. In contrast, \autoref{tab:generalization} shows that on \textit{VB-DemandEx}, compared to SEMamba, MambAttention yields a PESQ score and SI-SDR which is bigger by only 0.03, and 0.09, respectively. As all models in \autoref{tab:dns2020} have comparable parameter counts, and our MambAttention model also slightly outperforms the baselines across all reported metrics on the \textit{DNS 2020} dataset, we argue that this indicates that our MambAttention model scales more effectively with respect to dataset size. This suggests that our proposed MambAttention block is better suited for leveraging large, diverse training data for speech enhancement tasks.

\subsection{Training and Inference Efficiency}
In \autoref{tab:runtime}, we report training time in GPU hours on the \textit{VB-DemandEx} and \textit{DNS 2020} training sets and inference time in seconds on the \textit{VB-DemandEx}, \textit{DNS 2020} without reverb, and \textit{EARS-WHAM\_v2} test sets for our MambAttention model and the discriminative baselines trained in this paper. As NVIDIA GPUs are more common than AMD GPUs in published AI research, we report training and inference time on NVIDIA GPUs. The training times for each model are reported by training 1 seed on four NVIDIA L40S GPUs, and the inference time is computed by running inference on a single NVIDIA L40S GPU. In addition, we report FLOPs, which are computed based on processing a single \SI{2}{s} audio clip on one GPU.

\autoref{tab:runtime} shows that training our MambAttention model requires approximately \SI{13}{\%} more GPU hours than the Mamba baseline. For inference, we observe that the difference in inference time between the Mamba baseline and our MambAttention model becomes large on the \textit{EARS-WHAM\_v2} test set. We attribute this to the fact that the utterances in the \textit{EARS-WHAM\_v2} test set are considerably longer than the other test sets. Thus, the quadratic complexity of the MHA modules has a bigger impact on inference time. Our MambAttention model remains significantly more efficient that the Conformer- and xLSTM-based baselines with respect to both training and inference time. As shown in \autoref{tab:runtime}, our MambAttention model only requires slightly more FLOPs than the Mamba baseline, and it requires less FLOPs than the LSTM, xLSTM, and Conformer baselines.

Since the LSTM-based baseline does not contain any pre-up or post-down projections, as opposed to the xLSTM, Mamba, and MambAttention models, we hypothesize that this is the reason for lower training and inference time.

\section{DISCUSSION AND LIMITATIONS}\label{sec:discussion}
While our MambAttention model displays superior generalization performance for speech enhancement compared to discriminative LSTM, xLSTM, Mamba, and Conformer-based baselines, it does come at a cost. The addition of MHA adds additional trainable parameters, and due to using scaled dot-product attention, we no longer have linear scalability with respect to the input sequence length. This is one of the main advantages of newer sequence models such as Mamba \cite{gu2024mamba} and xLSTM \cite{xlstm}. Hence, \autoref{tab:runtime} shows an increase in training time by approximately \SI{13}{\percent} compared to the Mamba baseline.
However, as we train on \SI{2}{s} audio clips and run inference on at most \SI{10}{s} chunks of audio clips, the impact of the quadratic complexity of the MHA modules becomes less noticeable. Thus, this may only become an issue for real-time speech enhancement or for processing longer audio clips. To overcome the quadratic complexity of self-attention, recent works have introduced IO-aware exact attention algorithms and approximate methods resulting in significant reductions in runtime \cite{dao2023flashattention2, shah2024flashattention3}. These algorithms potentially counteract the computational downsides of using MHA. 

As observed in \autoref{tab:generalization}, all models perform similar when trained and evaluated on our proposed \textit{VB-DemandEx} dataset. This result is in line with our previous work in \cite{xlstm-senet}, where we reported that LSTM-, xLSTM-, Mamba-, and Conformer-based models perform similarly on \textit{VoiceBank+Demand} \cite{voicebank, demand}. Thus, our results add solid evidence to the conclusions drawn in \cite{xlstm-senet}, since our proposed \textit{VB-DemandEx} dataset features substantially lower SNRs and more noise types compared to \textit{VoiceBank+Demand}. The lack of performance differentiation between models, despite differences in noise diversity and SNRs across training datasets, raises concerns about solely using small-scale datasets for benchmarking speech enhancement performance. Thus, exclusively presenting performance on such datasets may obscure differences in model performance that only become apparent on larger and more diverse datasets or in mismatched speaker, noise, and recording conditions as observed in \autoref{tab:dns2020} and \autoref{tab:generalization}, respectively.
\section{CONCLUSION}\label{sec:conclusion}
In this paper, we proposed a novel MambAttention model that combines Mamba and shared multi-head attention for generalizable single-channel speech enhancement. To evaluate its performance, we introduced \textit{VB-DemandEx}, a new speech enhancement dataset based on \textit{VoiceBank+Demand} but with lower SNRs and more challenging noise types. When trained on \textit{VB-DemandEx}, our MambAttention model outperforms state-of-the-art discriminative LSTM-, xLSTM-, Mamba-, and Conformer-based systems of similar complexity across all reported metrics when evaluating on the two out-of-domain datasets: \textit{DNS 2020} without reverb and \textit{EARS-WHAM\_v2}, while matching their performance on the in-domain \textit{VB-DemandEx} dataset. Moreover, our MambAttention model matches or ourperforms generative diffusion models and is comparable with LM-based speech enhancement systems, further demonstrating its generalization performance. Detailed ablation studies reveal that the placement of the multi-head-attention modules significantly affect the generalization performance of our MambAttention model. Additionally, we found that the weight sharing mechanism positively affects generalization performance, while slightly reducing the overall parameter count. We also tested multi-head attention-augmented LSTM and xLSTM variants, which improved their generalization performance but remained inferior to our MambAttention model. Finally, results on the large-scale \textit{DNS 2020} dataset demonstrate that our MambAttention model scales more effectively with dataset size, achieving superior in-domain performance across all reported metrics compared to state-of-the-art LSTM-, xLSTM-, Mamba-, and Conformer-based baselines of similar complexity.

While our MambAttention model matches or outperforms state-of-the-art baselines in generalization performance, exploring real-time performance will be the focus of our future work. This will require an update to the entire MambAttention model, as neither the feature encoder, the time- and frequency-multi-head attention modules, the time- and frequency-Mamba blocks, nor the decoders are causal. However, we believe this would be a feasible research direction, since Mamba- and attention-based models have already shown potential for real-time speech enhancement \cite{groot2025cleanumamba, pandey2021dense}.

\bibliographystyle{IEEEtran}
\bibliography{mybib}
\vfill\pagebreak
\end{document}